\documentclass[pdflscape,lscape,iop,apj,numberedappendix,appendixfloats,stfloats,url,microtype,iop,revtex4]{emulateapj}

\usepackage{paralist,amsmath,amssymb}
\usepackage{longtable,rotating}
\usepackage{float}
\usepackage{morefloats}
\usepackage{placeins}
\usepackage{xfrac}
\usepackage{pdflscape}
\usepackage{afterpage}
\usepackage{capt-of}
\usepackage{lipsum}

    {\pagebreak[4]\global\pdfpageattr\expandafter{\the\pdfpageattr/Rotate 90}}%
    {\pagebreak[4]\global\pdfpageattr\expandafter{\the\pdfpageattr/Rotate 0}}%
\usepackage{threeparttable}

\graphicspath{{./}}

\shorttitle{Broadband radio spectropolarimetric variability in blazars}
\shortauthors{Anderson et al.}

\begin{document}

\title{Blazar jet evolution revealed by multi-epoch broadband radio polarimetry}


\author{
C. S. Anderson\altaffilmark{1}$^{\dagger}$,
S. P. O'Sullivan\altaffilmark{2},
G. H. Heald\altaffilmark{1},
T. Hodgson\altaffilmark{1,3},
A. Pasetto\altaffilmark{4},
B. M. Gaensler\altaffilmark{5}
}

\altaffiltext{$\dagger$}{{\bf craig.anderson@csiro.au}}
\altaffiltext{1}{CSIRO Astronomy \& Space Science, 26 Dick Perry Avenue, Kensington, WA 6151}
\altaffiltext{2}{Hamburger Sternwarte, Universit\"{a}t Hamburg, Gojenbergsweg 112, 21029 Hamburg, Germany}
\altaffiltext{3}{International Centre for Radio Astronomy Research - Curtin University, GPO Box U1987, Perth, WA 6845, Australia}
\altaffiltext{4}{Istituto de Radioastronom\'{i}a y Astrof\'{i}sica (IRyA-UNAM), Antigua Carretera a P\'{a}tzcuaro, Morelia, Michoac\'{a}n, Mexico}
\altaffiltext{5}{Dunlap Institute for Astronomy and Astrophysics, 50 St. George Street, Toronto, ON M5S 3H4, Canada}

\begin{abstract}

We investigate the previously proposed possibility that multi-epoch broadband polarimetry could act as a complement or limited proxy for VLBI observations of blazars, in that the number of polarised emission components in the jet, and some of their properties and those of the foreground environment, might be inferred from the object's time-varying 1D Faraday depth spectrum (FDS) alone. We report on a pilot-scale experiment designed to establish the basic plausibility and utility of this idea. We analyse temporal changes in the complex polarisation spectra of nine spatially unresolved (at arcsecond scales) blazars in two epochs separated by $\sim$5 years, using data taken with the Australia Telescope Compact Array. The data allow for precise modelling, and we demonstrate that all objects in our sample show changes in their polarisation spectrum that cannot be accounted for by uncertainties in calibration or observational effects. By associating polarised emission components across epochs, we infer changes in their number, intrinsic fractional polarisation, intrinsic polarisation angle, rotation measure, and depolarisation characteristics. We attribute these changes to evolution in the structure of the blazar jets, most likely located at distances of up to tens of parsecs from the central active galactic nuclei. Our results suggest that continued work in this area is warranted; in particular, it will be important to determine the frequency ranges and temporal cadence most useful for scientifically exploiting the effects.

\end{abstract}


\section{Introduction}\label{sec-intro}

\noindent Powerful jets launched by active galactic nuclei (AGN) drive evolution and ecology in the cosmos (e.g. \citealp{DiMatteo2005,MN2012}, \citealp{Harrison2018} and references therein). The processes that govern their formation, collimation, and propagation are not entirely understood, but depend on the detailed magnetoionic structure of the jets and their surroundings. 

From an observational point of view, progress in studying these jets has been driven both by very long baseline interferometry (VLBI) (e.g. \citealp{Udomprasert1997,Taylor2000,ZT2002,ZT2004,Jorstad2005,ZT2005,Ojha2010,Fuhrmann2016,Lister2016}, and references in each) and high-cadence single dish monitoring (e.g. \citealp{Dent1965,Aller1967,Aller1970a,Aller1970b,Komesaroff1984,Salonen1987,Steppe1988,Kovalev2002,Bach2007,Aller2017,Agudo2018}), with both techniques typically exploiting multi-frequency coverage over broad but sparsely-sampled frequency bands, often in full polarisation. They have proven highly complementary. In broad terms, they have revealed radio jets to be inhomogeneous sources of synchrotron radiation on sub-kpc scales --- often dominated by a handful of discrete emission components associated with shocks --- that are characterised by (among other properties) their respective fractional linear polarisation, intrinsic polarisation angle, Faraday rotation measure (RM), and Faraday depth dispersion. In general, these observables vary throughout the jet in characteristic ways, and are used to assess models of the geometric configuration of these regions and the physical conditions existing therein. Blazars (a class consisting of both flat spectrum radio quasars, or FSRQ, and BL Lac objects) observed at scales below tens of parsecs are particularly interesting in this regard, because aided by relativistic effects (e.g. \citealp{LV1999,Britzen2008,Lister2016,Saikia2016}), we observe their structure evolve on human timescales. This fundamentally helps us piece together the physics occuring therein. For example, the time-evolution of low-spatial-resolution polarisation data furnished key evidence for the foundational shock-in-jet model of blazar emission \citep{MG1985,Valtaoja1992,Spada2001,Turler2011} as well as its alternatives, including time-variable doppler beaming (e.g. \citealp{CK1992}), the turbulent extreme multi-zone model \citep{Marscher2017}, and magnetic reconnection models (e.g. \citealp{Petropoulou2016,Morris2018,Zhang2018}). Moreover, changes in RM and polarisation structure measured using VLBI have helped isolate the physical properties of jet components from that of their surrounding environments \citep{Gomez2000,ZT2001,ZT2004,MD2005,Asada2008,Gomez2008,Mahmud2009,Gomez2011,Hovatta2012,Lico2017}, and to study the kinematics and dynamics of the jet components themselves \citep{Homan2009,Homan2015,Lister2009,Lister2013}.

Nevertheless, observational challenges persist. In particular, for the frequency bandwidths and baselines available to contemporary VLBI arrays, key structures of interest in and around the jets either cannot be resolved, or are resolved out (see e.g. \citealp{TZ2010,BM2010,Kharb2010,Pudritz2012}). Moreover, the logistics of obtaining coeval data in multiple frequency bands --- required for calculating reliable RMs --- are often prohibitive for large samples. Finally, sparse frequency coverage renders the interpretation of RM measurements difficult, in an analogous manner to how sparse $uv$ coverage affects the interpretation of aperture synthesis images (see e.g. \citealp{BdB2005,Anderson2016b} for further explanation). All of this will remain the case for the foreseeable future; novel observational approaches in this area are therefore desirable. \\

One such approach --- broadband spectropolarimetry --- seeks to exploit frequency-dependent interference effects in polarised emission to constrain the magneto-ionised structure of sources on scales below the observing resolution (e.g. \citealp{Burn1966,Conway1974,Tribble1991,Sokoloff1998,Law2011,OSullivan2012,Gaensler2015,Anderson2016}). The way in which this information is encoded can be described as follows: Linear polarisation states are represented by a complex vector $\boldsymbol{P}$, related to the Stokes parameters $Q$ \& $U$, the polarization angle $\psi$, the fractional polarization $p$, and the total intensity $I$ as:
 
 \begin{equation}
\boldsymbol{P} = Q + iU = pIe^{2i\psi}
\label{eq:ComplexPolVec}
 \end{equation}

\noindent In transiting magnetized plasma between a point $L$ and the observer, linearly polarized radiation will be Faraday-rotated by an amount equal to

\begin{equation}
\Delta\psi= \phi\lambda^2
\label{eq:rotation}
\end{equation}

\noindent where $\psi$ is the polarization angle, $\lambda$ is the observing wavelength, and $\phi$ is the Faraday depth, given by  

 \begin{equation}
\text{$\phi$}(L) = 0.812 \int_{L}^{\text{telescope}} n_e\boldsymbol{B}\cdot\text{d}\boldsymbol{s}~\text{rad m}^{-2}
\label{eq:FaradayDepth}
 \end{equation}
 
\noindent and, in turn, $n_e$ [cm$^{-3}$] \& $\boldsymbol{B}$ [$\mu$G] are the thermal electron density and magnetic field along the LOS respectively. The net observable polarization $\boldsymbol{P}(\lambda^2)$ is obtained by summing the polarized emission from all possible Faraday depths within the synthesized telescope beam:

 \begin{equation}
\boldsymbol{P}(\lambda^2) = \int_{-\infty}^{\infty} \boldsymbol{F}(\phi) e^{2i\phi\lambda^2} d\phi
\label{eq:SumPol}
 \end{equation}
 
\noindent The function $\boldsymbol{F}(\phi)$ (which we call the Faraday spectrum) specifies the distribution of polarized emission as a function of Faraday depth along the LOS, and possesses units of Jy rad$^{-1}$ m$^2$ sr$^{-1}$ for a source which is extended in the plane of the sky, and extended in Faraday depth. 

Burn's theory raises the prospect of isolating and studying polarised emission structure in radio jets \emph{spectrally} rather than spatially. Building on this, \citet{Law2011}, \citet{OSullivan2012}, and \citet{Anderson2016} (hereafter L11, OS12, A16 respectively) pointed out that:

\begin{itemize}
\item the properties of multiple spatially-unresolved emission components can indeed be disentangled (at least under certain circumstances) using modern broadband polarimetric analysis (e.g. \citealp{Anderson2015,Anderson2016,Anderson2018,Kim2016,Pasetto2016,Pasetto2018,Farnes2017,OSullivan2017,OSullivan2018,Vernstrom2017}). 
\item the polarimetric interference effects observed therein must be generated by sub-kpc-scale, and possibly by parsec-scale, emission and Faraday rotation structure in and around the sources themselves
\item the parsec-scale polarisation structure of blazars are known to change on month- to year-long timescales, as components brighten, fade, change their internal magnetic configuration, and propagate down the jet, back-illuminating different structures in the foreground as they do
\end{itemize}

\noindent It follows, suggest L11, OS12, and A16, that multi-epoch broadband polarimetric analysis could act as a complement or limited proxy for very long baseline interferometry (VLBI), insofar as it might be possible to identify the number of dominant polarised emission components in the jets, infer certain of their physical properties and those of the foreground environment, and track changes in the configuration of the system as it evolves.\\ 

In this work, we aim to assess whether the relevant time-dependent polarimetric interference effects are observable in practice, and are likely to be useful as a tracer of structural evolution in blazar jets. We report the results of a pilot experiment in which nine sources have been observed in two epochs separated by $\sim$5 years using the Australia Telescope Compact Array (ATCA) over densely-sampled multi-GHz bands in full polarisation. The paper is set out as follows. We describe the sample in Section \ref{sec-sample}, and our observations, calibration, and imaging in Section \ref{sec-obs}. Our analysis, results, and discussion are presented in Sections \ref{sec-analysis}, \ref{sec-results}, and \ref{sec-discussion} respectively.

\section{Sample construction}\label{sec-sample}

\noindent We selected sources from the OS12 and A16 samples (hereafter referred to as the OS12-selected sources and A16-selected sources) that \emph{must} be dominated by sub-kpc-scale emission structure (based on upper limits to their angular extent provided by these authors), and are \emph{likely} to be dominated by pc-scale emission structure, based on a variety of different arguments supplied by OS12, A16, and references in each (also see Section \ref{sec-bfmodelcomparo}). These general selection criteria were adopted to maximise the chance of observing the sought-after temporal polarisation changes in this small pilot study. The final sample includes all four sources from OS12, and five of the brightest, spatially-unresolved radio sources from A16. The sources are listed in Table \ref{tab:SourceDat}, alongside some of their key properties. We note here that the source PKS B0515-674 was observed as a calibrator for this project, and is included in our analysis as a control.

\begin{deluxetable*}{lcllccccc} 
\tabletypesize{\footnotesize} 
\setlength{\tabcolsep}{0.015in} 
\tablewidth{0pt} 
\tablecaption{Selected attributes of the sample sources} 
\\ 
\tablehead{
\colhead{(1)} & \colhead{(2)} & \colhead{(3)} & \colhead{(4)} & \colhead{(5)} & \colhead{(6)} & \colhead{(7)}  & \colhead{(8)}  & \colhead{(9)} 
\\ 
\colhead{Source} & \colhead{Archival} & \colhead{RA} & \colhead{Decl.} & \colhead{Common} & \colhead{$I_{1.4}$} & \colhead{$\alpha_{1.4}$} & \colhead{$b$} & \colhead{H$\alpha$} 
\\ 
\colhead{} & \colhead{designation} & \colhead{(J2000)} & \colhead{(J2000)} & \colhead{freq. bands} & \colhead{(2012 epoch)} & \colhead{(2012 epoch)} 
\\ 
\colhead{} & \colhead{} & \colhead{} & \colhead{} & \colhead{[cm]}  & \colhead{[Jy]} & \colhead{}  & \colhead{[deg]} & \colhead{[R]} 
}
\startdata 
PKS B0454-810 & & 04:50:05.4 & -81:01:02.2 & 16 & 0.81 & +0.45 & -31.4 & 0.9 \\
PKS B0515-674* & & 05:15:37.56 & -67:21:27.78 & 16, 6, 3 & 1.54 & -0.76 & -34.1 & 18 \\
PKS B0517-726 & lmc\_s13  & 05:16:36.84 & -72:37:10.86 & 16, 6, 3  & 0.20 & +0.36 & -32.8 & 2.2 \\
PKS B0543-735 & lmc\_s11 & 05:41:50.69 & -73:32:14.01 & 16, 6, 3  & 0.99 & -0.42 & -30.8 & 1.1 \\
PKS B0545-649 & lmc\_c04 & 05:45:54.59 & -64:53:28.2 & 16, 6, 3  & 0.45 & -0.55 & -31.3 & 0.1 \\
PKS B0611-74 & lmc\_c03 & 06:10:12.9 & -74:32:05.95 & 16, 6, 3 & 0.73 & -0.83 & -28.8 & 0.6 \\
PKS B1039-47 & & 10:41:44.6 & -47:40:00.1 & 16 & 1.88 & -0.32 & +9.7 & 20 \\
PKS B1610-771 & & 16:17:49.2 & -77:17:18.5 & 16 & 3.54 & -0.36 & -18.9 & 3.0 \\
PKS B1903-802 & & 19:12:40.0 & -80:10:05.9 & 16 & 1.12 & -0.04 & -27.6 & 1.4
\enddata 
\tablecomments{*PKS B0515-674 is an unpolarised source that was observed as a check on our calibration. It is included in this work as a control. The entries in column 2 refer to a naming system employed by A16. Both the $\alpha_{1.4}$ (defined as $I\propto\nu^{-\alpha}$) and $I_{1.4}$ values (Columns 6 \& 7) have been quoted to 2 decimal places, which approximately reflect the limits placed on the variability of our flux calibrator PKS B1934-638 over a moving five year window by \citet{Tingay2003}. The formal fitting uncertainties are considerably lower. Column 8 lists the Galactic latitude of the source (1 dec. pl.), and Column 9 lists the H$\alpha$ intensity in Rayleighs at the position of each source (2 sig. figs.), extracted from the map presented by \citet{Finkbeiner2003}.} 
\label{tab:SourceDat} 
\end{deluxetable*}

\section{Observations, Calibration and imaging}\label{sec-obs}

\noindent We re-observed the sample sources using the Australia Telescope Compact Array (ATCA) in 1 MHz channelised continuum over 1.1--3.1 GHz, 4.5--6.5 GHz, and 8.0--10.0 GHz in full polarisation. We flagged one hundred 1 MHz channels at the bandpass edges because of a roll-off in sensitivity, resulting in an effective frequency coverage of 1.2--3.0 GHz (reduced to 1.3--3.0 GHz in this band after flagging radio frequency interference), 4.6--6.4 GHz, and 8.1--9.9 GHz, or 0.010--0.053 m$^2$, 0.0022--0.0042 m$^2$, and 0.00092--0.00137 m$^2$ in $\lambda^2$ space. Henceforth, we refer to these as the 16 cm, 6 cm, and 3 cm bands, respectively. We observed using a 6 km array configuration (6A) to maximise spatial resolution across all frequencies ($\sim$5" at 3 GHz, $\sim$1" at 9.9 GHz), and to match the observational setup of the archival observations. With one exception (described at the end of this section), each source was observed directly on-axis over a full 12 hr synthesis, for a total integration time of 2 hours summed over the frequency bands and split between 10 $uv$ cuts covering $\sim180$ degrees in parallactic angle. Details of our new observations, and those of the archival observations, are recorded in Table \ref{table:observations}.

\begin{deluxetable*}{cclccccc} 
\tabletypesize{\footnotesize} 
\tablecolumns{8} 
\tablewidth{0pt} 
\tablecaption{Summary of the observations used in this work} 
\tablehead{ 
\colhead{(1)} & \colhead{(2)} & \colhead{(3)} & \colhead{(4)} & \colhead{(5)} & \colhead{(6)} & \colhead{(7)} & \colhead{(8)} \\ 
\colhead{Observations} & \colhead{$\nu$ span} & \colhead{Epoch(s)} & \colhead{ATCA config.} & \colhead{Beam} & \colhead{$\Delta\phi$} & \colhead{$\phi_\text{Max-scale}$} & \colhead{$|\phi_\text{Max}|$}\\ 
\colhead{} & \colhead{[GHz]} & \colhead{} & \colhead{} & \colhead{["]} & \colhead{[rad m$^{-2}$]} & \colhead{[rad m$^{-2}$]} & \colhead{[rad m$^{-2}$]}
} 
\startdata 
O'Sullivan+ (2012) & 1.21--3.10 & 2011 Jan 9, 20 & 6A & 5$\times$5 & 60 & 340 & 13000 \\
 Anderson+ (2016) & 1.35--9.90 & 2012 Feb 10--12 & 6A & 15$\times$15--1$\times$1 & 71 & 3400 & 7900  \\
  &  & 2012 Jun 22--24 & 6D &  &  &  &  \\
  &  & 2012 Aug 17--19 & 6A &  &  &  & \\
This work & 1.30--9.90 & 2017 Apr 13--15, 22 & 6A & 15$\times$15--1$\times$1 & 66 & 3400 & 7900 \\
This work (PKS B1039-47) & 1.3--3.10 & 2016 Jun 6 & 1.5B & 5$\times$5 & 79 & 340 & 7900
\enddata 
\tablecomments{The dissimilar low frequency limits quoted in column 2 are caused by differences in flagging strategy and time-variability of the radio frequency interference (RFI) environment. The quoted range in synthesized beam size (column 5) corresponds to the change in such between 1.3 GHz and the upper frequency limit of each set of observations.} 
\label{table:observations}
\end{deluxetable*}

Detailed descriptions of the (standard) calibration and imaging procedures adopted for the 2012 data are provided by OS12 and A16. We outline their approach here, since our new data were processed in the same way. Radio frequency interference was flagged throughout the calibration process using the {\sc sumthreshold} algorithm \citep{Offringa2010}. Data below 1.3 GHz, and between 2.68 and 3.1 GHz, were flagged completely. The middle period of our 16cm band observations were also severely affected by so-called `mid-week' RFI \footnote{See notes in the ATCA manual --- http://www.narrabri.atnf.csiro.au/observing/users\_guide/html/ atug.html}. The main effect of this has been to reduce our sensitivity compared to the archival observations. Daily observations of PKS B0823-500 were used to calibrate the bandpass and flux scale. This source has a year-on-year flux variability of up to 10\%, meaning our absolute flux scale is uncertain by the same amount\footnote{See notes on this source in the ATCA calibrator database --- http://www.narrabri.atnf.csiro.au/calibrators/calibrator\_ database.html}. This does not affect our primary analysis, because it was conducted on quotients of polarisation products that are tied to the same flux scale (see the next paragraph). The time-dependent complex antenna gains and on-axis polarization leakage were calibrated using PKS B0515-674, and in the case of sources selected from OS12, the target PKS B0611-74. We verified the integrity of the calibration results by deriving the complex gains and leakage from other sources in our sample, and cross-applying them. The gains, leakages, and spectra of the sources were found to be effectively indistinguishable in all cases. The post-calibration on-axis polarisation leakage is typically $\sim0.05$\% of Stokes I (see \citealp{Schnitzeler2011, Anderson2016}). The data were then self-calibrated in both phase (two rounds) and amplitude (one round), with solutions derived in 128 MHz sub-bands then interpolated and applied continuously across the full band. 

We imaged the sources in Stokes $I$, $Q$, $U$, and $V$ at 20 MHz intervals through each of the 16 cm, 6 cm, and 3 cm bands, using robust = 2 weighting \citep{Briggs1995} to achieve maximum sensitivity. The full-Stokes flux densities were taken to be the pixel value at the location of the Stokes $I$ maxima of each source in the respective Stokes $I$, $Q$, $U$, and $V$ images. The corresponding uncertainties were estimated as the root-mean-square value of an adjacent source-free region in the maps. We conducted our spectropolarimetric analysis on the fractional Stokes parameters $q=Q/I$ and $u=U/I$, which in this case were obtained by fitting a polynomial model to log($I$) versus log($\nu$) (e.g. \citealp{SH2012}), calculating the corresponding $I(\lambda^2)$ model, and dividing this out of Stokes $Q(\lambda^2)$ and $U(\lambda^2)$. 

As previously indicated, there was one exception to this workflow: the source PKS B1039-47. Due to a coordinate input error, it was not observed during the 2017 observations. Instead, we obtained data for this source from the C007 project, which monitors ATCA secondary calibrators. The source was observed for this project in June 2016 in the 16cm band only (see Table \ref{table:observations}). Its Stokes $I$, $Q$, $U$, and $V$ values were extracted directly from measured visibilities at 20 MHz intervals. The polarisation leakages were derived from the primary calibrator PKS B1934-638, which is accurate to 0.1\%. The absolute polarisation angle is calibrated using an injected signal from a noise source, and is accurate to 1\%. Thus, the data are of sufficient quality to use in this project.

\section{Analysis}\label{sec-analysis}

\subsection{Spectropolarimetric modelling}\label{sec-modelling}

\noindent OS12 and A16 used dissimilar techniques for their spectropolarimetric analysis. We therefore re-analysed all of our data consistently as follows. We modelled the frequency-dependence of individual polarised emission components as \citep{OSullivan2017}:

\begin{multline}
\boldsymbol{P}_j(\lambda^2) = p_{0[j]}e^{2i({\psi_{0[j]}}+\text{RM}_{[j]}\lambda^2)} \\ \times e^{-2\sigma_{\text{RM}[j]}^2\lambda^4} \times \text{sinc}(\Delta \phi_{[j]}\lambda^2)
\label{eqn:ShaneMod}
\end{multline}

\noindent where $p_{0[j]}$, $\psi_{0[j]}$, and RM$_{[j]}$ are the initial fractional polarisation, initial polarisation angle, and rotation measure of the $j$th emission component (respectively), and $\sigma_{\text{RM}[j]}$ and $\Delta \phi_{[j]}$ parameterise Faraday-dispersive effects. The first exponential term in Eqn. \ref{eqn:ShaneMod} models Faraday rotation, the second such term models depolarization by an external turbulent magnetoionic foreground \citep{Burn1966}, while the last term can model depolarisation from a number of possible sources \citep{Schnitzeler2015}, including that from mixed synchrotron-emitting and Faraday-rotating plasma, or RM gradients across emitting components, as particular cases of interest.  We note that we attempt to model the polarisation spectrum of each source in terms of Faraday rotation, Faraday depolarisation, and multicomponent interference effects (e.g. \citealp{GR1984}) only --- we neglect effects related to optical depth (e.g. see \citealp{Aller1970b}) and spectral index (e.g. see \citealp{Burn1966}). We discuss this further in Section \ref{sec-veracity}. 

We constructed 34 unique \emph{model types} to fit to each source --- that is, model types incorporating either one, two, or three emission components, whose frequency-dependent polarisation behaviour is described by one of the following variants of Eqn. \ref{eqn:ShaneMod}: (1) all of the terms can take non-zero values, (2) the $\sigma_{\text{RM}}$ term is set to zero, (3) the $\Delta\phi$ term is set to zero, or (4) both the $\sigma_{\text{RM}}$ and $\Delta\phi$ terms are set to zero. These model types describe the combined action of external Faraday rotation + internal Faraday dispersion + foreground RM dispersion, external Faraday rotation + internal Faraday dispersion, external Faraday rotation + foreground RM dispersion, or external Faraday rotation acting alone. We refer to such components with the designations `M', `I', `E', and `R' respectively, in reference to mixed (internal/external Faraday dispersive effects), internal (Faraday dispersion), external (differential Faraday rotation), and rotation (only). We refer to different model types by chaining these letter designations together --- for example, an RIE model type incorporates three emission components, possessing the aforementioned depolarisation behaviours. The ordering of the letters is unimportant. 

\subsection{Fitting}\label{sec-fitting}

We estimated the best-fit parameters for our different model types using Bayesian methods. The posterior probability $\text{P}(\boldsymbol{\theta}|\boldsymbol{d},\text{M})$ for a vector of model parameters ($\boldsymbol{\theta}$), given a vector of data ($\boldsymbol{d}$) and a model type (M), can be calculated using Bayes' theorem:

\begin{equation}
\text{P}(\boldsymbol{\theta}|\boldsymbol{d},\text{M}) = \frac{\text{P}(\boldsymbol{d}|\boldsymbol{\theta},\text{M}) \times \text{P}(\boldsymbol{\theta}|\text{M})}{\text{P}(\boldsymbol{d}|\text{M})} 
\label{eq:bayes}
\end{equation}
 
\noindent For polarization data ($q_i$,$u_i$) and a model (q$_{mod,i}$,u$_{mod,i}$), the likelihood $\text{P}(\boldsymbol{d}|\boldsymbol{\theta},\text{M})$ is:

\begin{multline}
\text{P}(\boldsymbol{d}|\boldsymbol{\theta},\text{M}) = \prod\limits_{i=1}^n \frac{1}{\pi\sigma_{q_{i}}\sigma_{u_{i}}} \\ \times \text{exp}\Bigg(-\frac{(q_i-q_{mod,i})^2}{2\sigma_{q_{i}}^2}-\frac{(u_i-u_{mod,i})^2}{2\sigma_{u_{i}}^2}\Bigg)
\label{eq:likelihood}
\end{multline}

\noindent For $\text{P}(\boldsymbol{\theta}|\text{M})$ --- i.e. the prior degree of belief that the model parameters will assume a given set of values --- we used a uniform probability density function (PDF) in ranges where these values were physically acceptable and constrained by the data. That is, when data was available in the 16 cm, 6 cm and 3 cm bands (see below), we used: [$-\pi/2$,$\pi/2$) rad for $\psi_{0[j]}$, [0,0.7] for $p_{0[j]}$, [-2000,2000] for $\text{RM}_{[j]}$, and [0,2000] rad m$^{-2}$ for both $\sigma_{\text{RM}[j]}$ and $\Delta \phi_{[j]}$. When only the 16 cm band was available, we modified the $\sigma_{\text{RM}[j]}$ and $\Delta \phi_{[j]}$ ranges to be [0,85] rad m$^{-2}$ and [0,85] rad m$^{-2}$ respectively, since the depolarisation induced by larger values of such at $\lambda^2>0.01$ m$^2$ means the parameters are poorly constrained. The prior PDF was set to zero outside these ranges. 

To sample the posterior PDF and determine the best-fit model parameters for each model type, we used {\sc pymultinest} \citep{Buchner2014} --- a {\sc python} interface to the {\sc multinest} package \citep{Feroz2009,Feroz2013}, which is an implementation of the nested sampling algorithm \citep{Skilling2004}. {\sc multinest} is an efficient tool for sampling potentially multi-modal and relatively high-dimensional ($n<30$) posterior PDFs. Since components of the same type are degenerate in the fitting process (e.g. for an RRR model), interchange modes can appear in the posterior PDF (known as the label swapping problem), manifesting as a bi- (or tri-) modal marginalised posterior. We manually identified and eliminated samples belonging to the duplicate modes. From the resulting marginal posterior PDFs, we take the median, and whichever is smaller of the 16th and 84th percentiles, to represent the best fit parameter values and their characteristic uncertainties (respectively) for a given model type. The full marginal posteriors for the best-fit model types (see Section \ref{sec-comparison}) are provided in Appendix \ref{sec-appendA} (available in the online version of this paper); they are mostly close to Gaussian, and the 16th and 84th percentiles are typically close in value, meaning they provide a reasonable parameterisation of the fitting uncertainty.

OS12 did not observe their target sources in the 6 cm and 3 cm bands. Thus, while we have observed all sources over 1.3--9.9 GHz band in the 2017 epoch, we only consider frequency ranges observed in both epochs for our main analysis, consistent with our primary goal of searching for and characterising spectral change. 

\subsection{Model type comparison}\label{sec-comparison}

To evaluate the best overall model type for a given source and epoch, we calculate the Bayes factor (assuming that there is no \emph{a priori} support for particular model types) as:

\begin{equation}
\text{B}_{\text{M}_0,\text{M}_1} = \frac{\text{P}(\boldsymbol{d}|\text{M}_0)}{\text{P}(\boldsymbol{d}|\text{M}_1)}
\label{eq:bayesfactor}
\end{equation}

\noindent where M$_0$ and M$_1$ are two different model types, and

\begin{equation}
\text{P}(\boldsymbol{d}|\text{M}) = \int d\boldsymbol{\theta} ~\text{P}(\boldsymbol{d}|\boldsymbol{\theta},\text{M})\text{P}(\boldsymbol{\theta}|\text{M})
\label{eq:evidence}
\end{equation}

\noindent The integral in Eqn. \ref{eq:evidence} penalises model types that (a) cannot achieve a close fit to the data, or (b) open up excessive regions of parameter space to achieve a good fit, by comparison with other more efficient models. The value of the integral is robustly computed by {\sc multinest}. We recorded it for each of the 34 model types fitted to each polarisation spectrum, then computed the Bayes factor for all model type pairings. According to \citet{jeffreys1998}, $\text{M}_0$ is strongly favoured over $\text{M}_1$ when $\text{B}_{\text{M}_0,\text{M}_1} > 30$ --- a criterion we also adopt.

Thus, for both models and model types, we henceforth distinguish between those that achieve the best overall fit, those that represent plausible alternatives, and those which are strongly unfavoured (i.e. those models/types with the highest value of $\text{P}(\boldsymbol{d}|\text{M})$, those with $\text{B}_{M_{\text{best-fit}},M_1}<30$, and those with $\text{B}_{M_{\text{best-fit}},M_1}\geq30$, respectively).

We note that in principle, the accuracy of these methods could be tested using Monte Carlo techniques: An array of simulated sources could be generated using a specified number of model components with known spectral characteristics, and then corrupted with noise. Our methods (and others) could then be applied to determine the accuracy with which these properties can be extracted. While the value of such analysis is clear (see \citealt{Sun2015}), the planning required to appropriately bound the problem and execute the analysis in practice is substantial, and is beyond the scope of this work. We suggest that such a study might best be performed as a community data challenge, in the same style as \citet{Sun2015}, and as a direct successor to it.

\section{Results}\label{sec-results}

\subsection{Data and best-fit models for the 2012 and 2017 epochs}\label{sec-bfmodelsselect}

\begin{figure*}[htpb]
\centering
\includegraphics[width=0.8\textwidth]{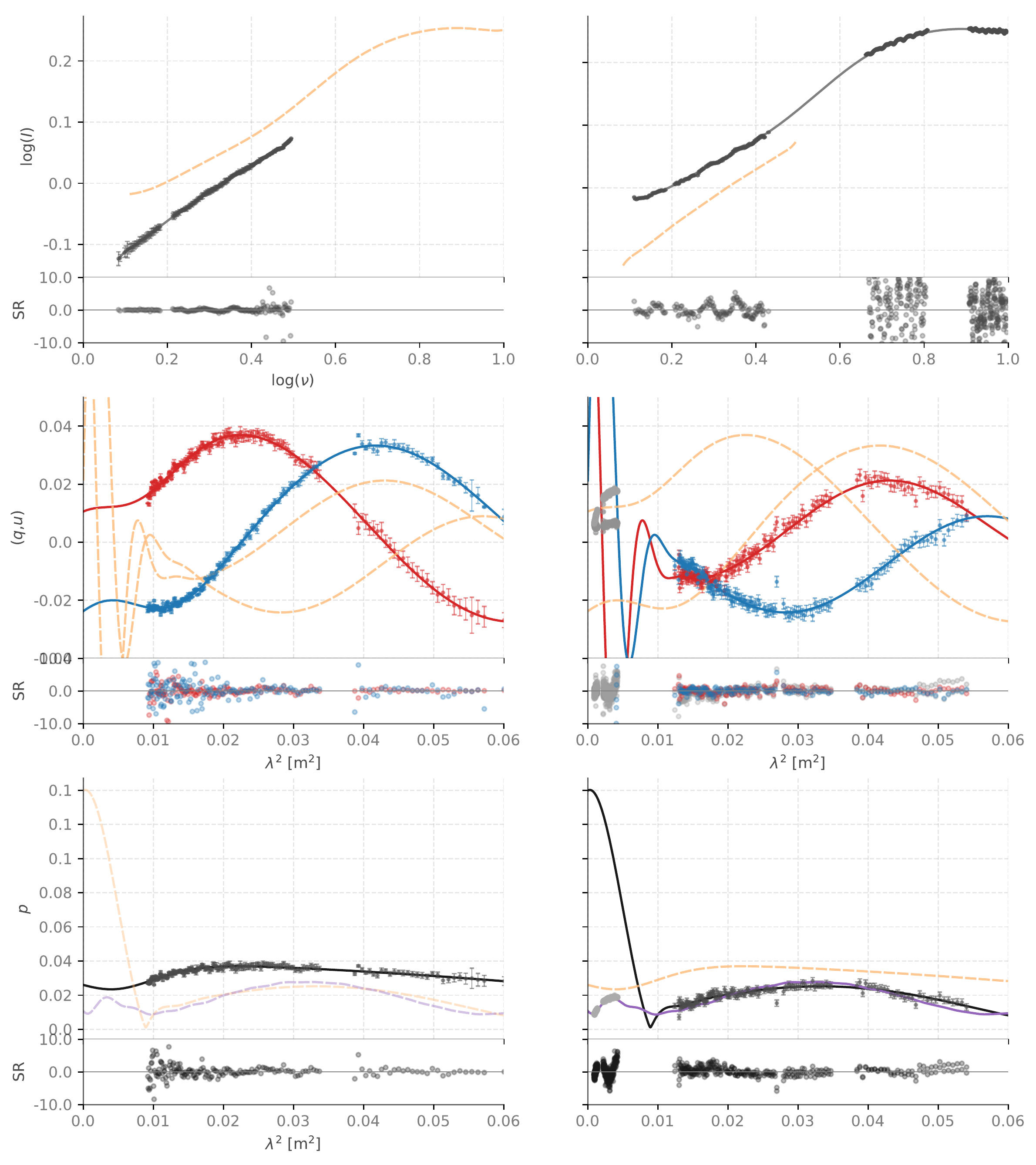}
\caption{Stokes $I$ (top row of panels), Stokes ($q$,$u$) data (middle row), and fractional polarisation ($p$) data for PKS B0454-810 in the 2012 and 2017 epochs (left and right column of panels, respectively). The main axes for each panel present the spectral data --- black points for Stokes $I$, red and blue points for Stokes $q$ and $u$ respectively, and black points for $p$. Light gray points for ($q,u$) and $p$ data indicate frequency ranges that were not observed in the 2012 epoch for the OS12 observations, and are thus ignored for the sake of our cross-epoch component association analysis. The error bars are derived from measurements of the root-mean-squared noise level in our Stokes $I$, $q$, and $u$ image cubes on a per-channel basis in regions adjacent to the targets. The best-fit models for the data are plotted in matching colours. To facilitate comparison between epochs, the best-fit model for both the 2012 and 2017 epochs are mirrored on the opposite epoch axes as dashed orange lines for each of the Stokes $I$, ($q,u$), and $p$ plots. For the OS12 sources, we indicate the best-fit model to the full frequency band in the 2017 epoch with a purple line, and mirror this on the 2012 epoch axis with a dashed purple line. We omit the Stokes ($q,u$) equivalent for the sake of clarity of the plots. The sub-axes on each panel are the standardised residuals (measured in standard deviations) for the difference between the data and the best-fit model. The standardised residuals for the 2017 epoch Stokes $I$ data exhibit systematic sinusoidal oscillations, which are residual errors from the bandpass calibration, as discussed in the main text.}
\label{fig:0454_comparo}
\end{figure*}

\begin{figure*}[htpb]
\centering
\includegraphics[width=0.8\textwidth]{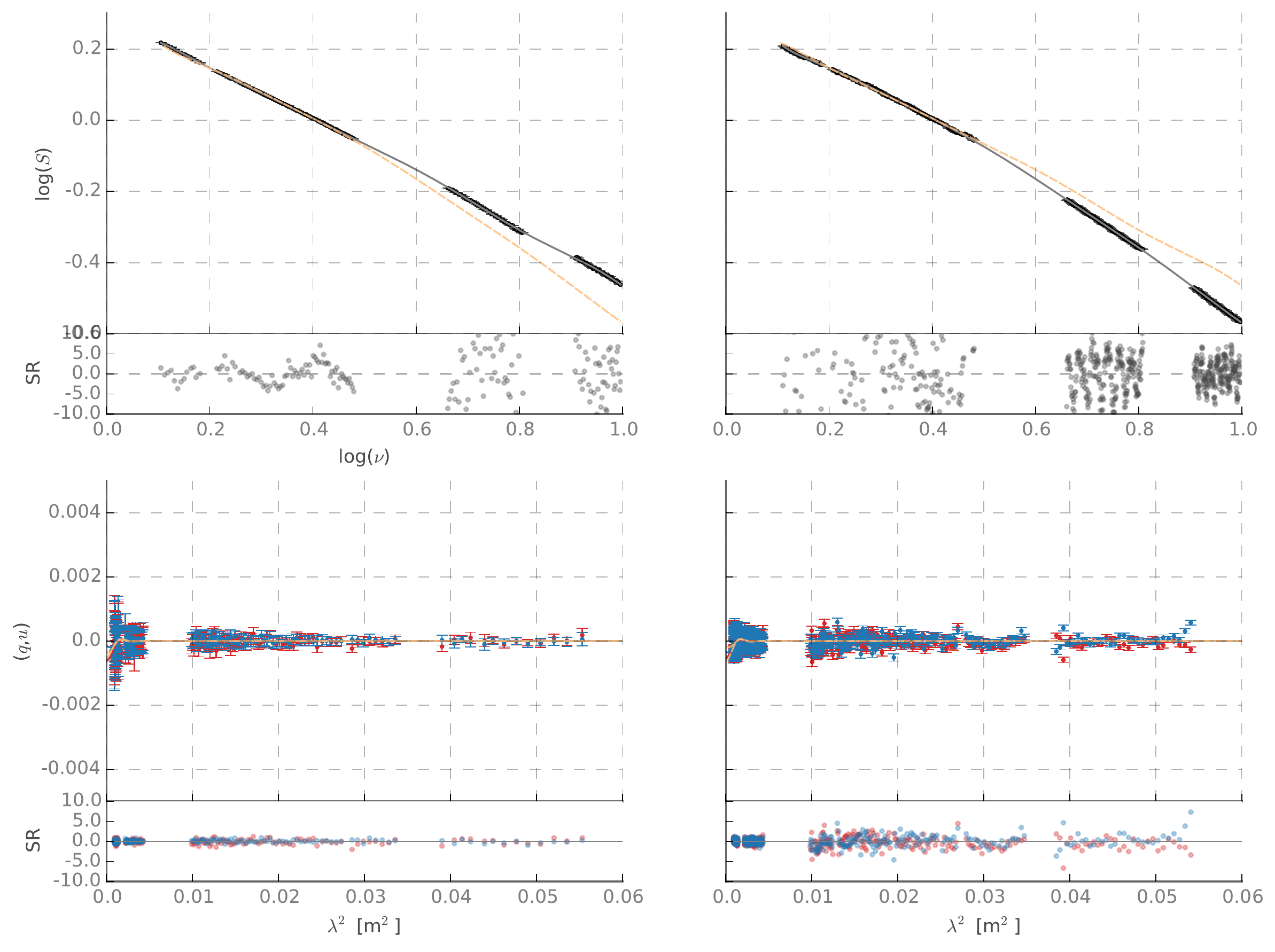}
\caption{As for Figure \ref{fig:0454_comparo}, but for PKS B0515-674.}
\label{fig:0515-674_comparo}
\end{figure*}

\begin{figure*}[htpb]
\centering
\includegraphics[width=0.8\textwidth]{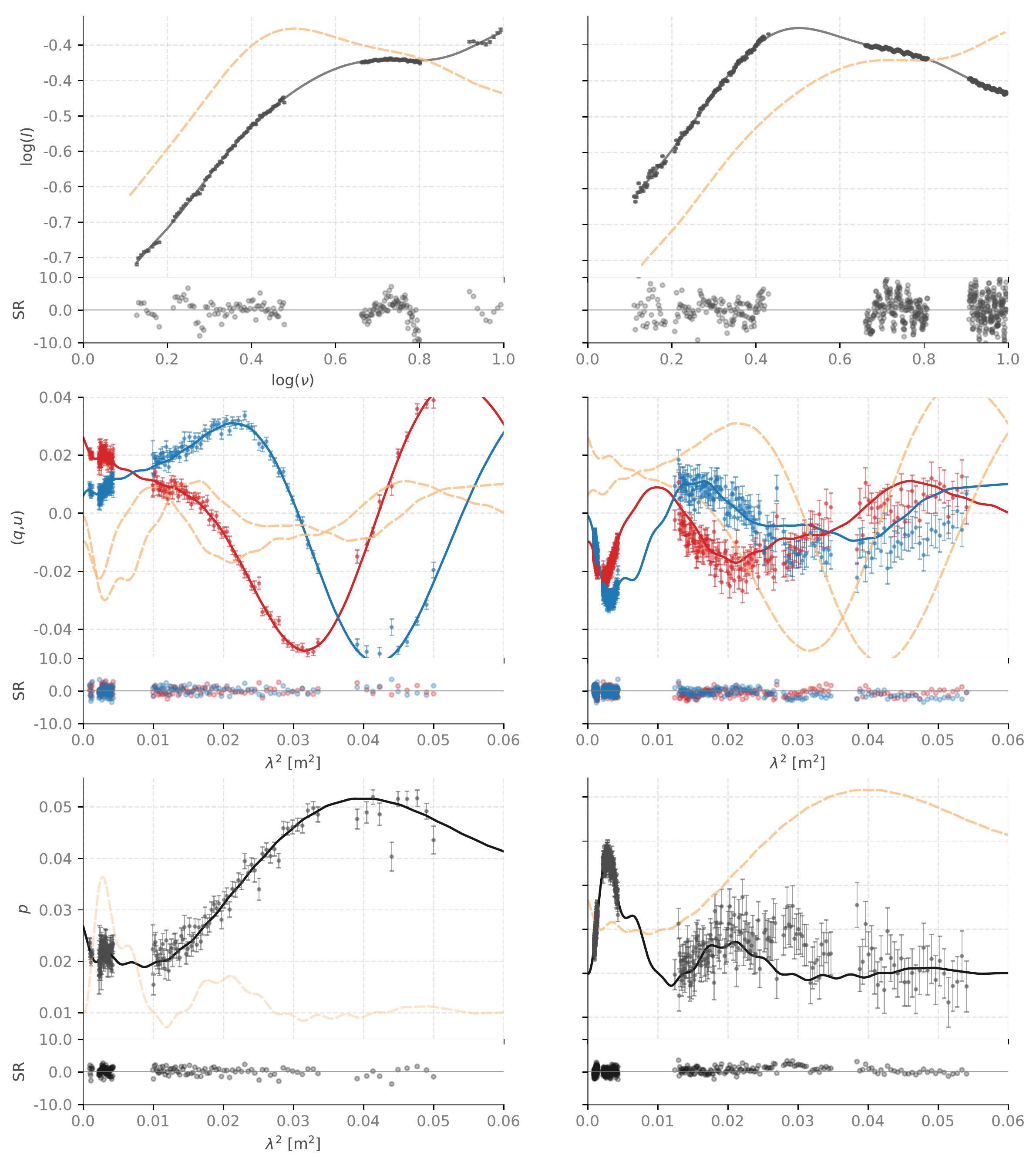}
\caption{As for Figure \ref{fig:0454_comparo}, but for PKS B0517-726.}
\label{fig:13_comparo}
\end{figure*}

\begin{figure*}[htpb]
\centering
\includegraphics[width=0.8\textwidth]{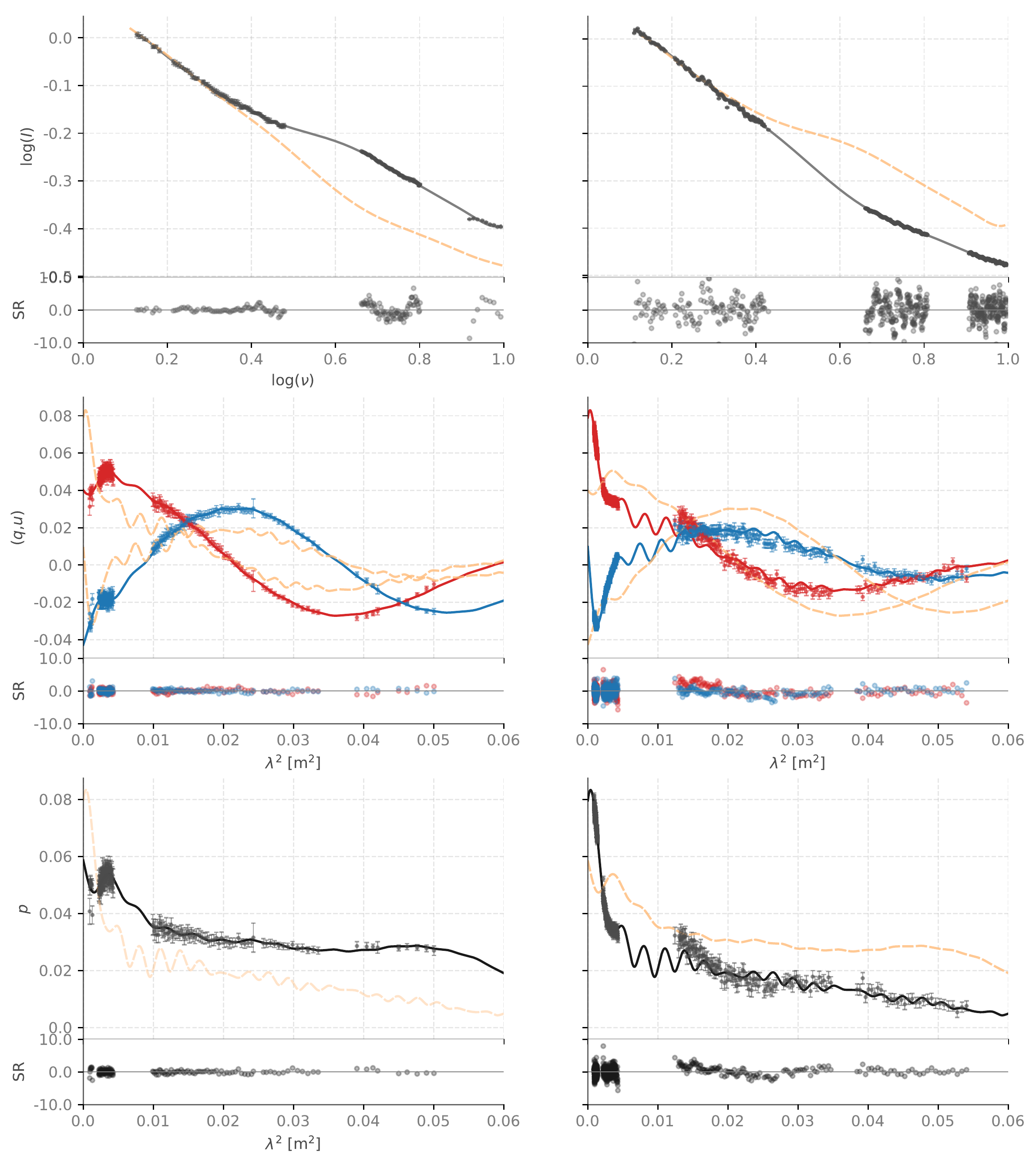}
\caption{As for Figure \ref{fig:0454_comparo}, but for PKS B0543-735.}
\label{fig:11_comparo}
\end{figure*}

\begin{figure*}[htpb]
\centering
\includegraphics[width=0.8\textwidth]{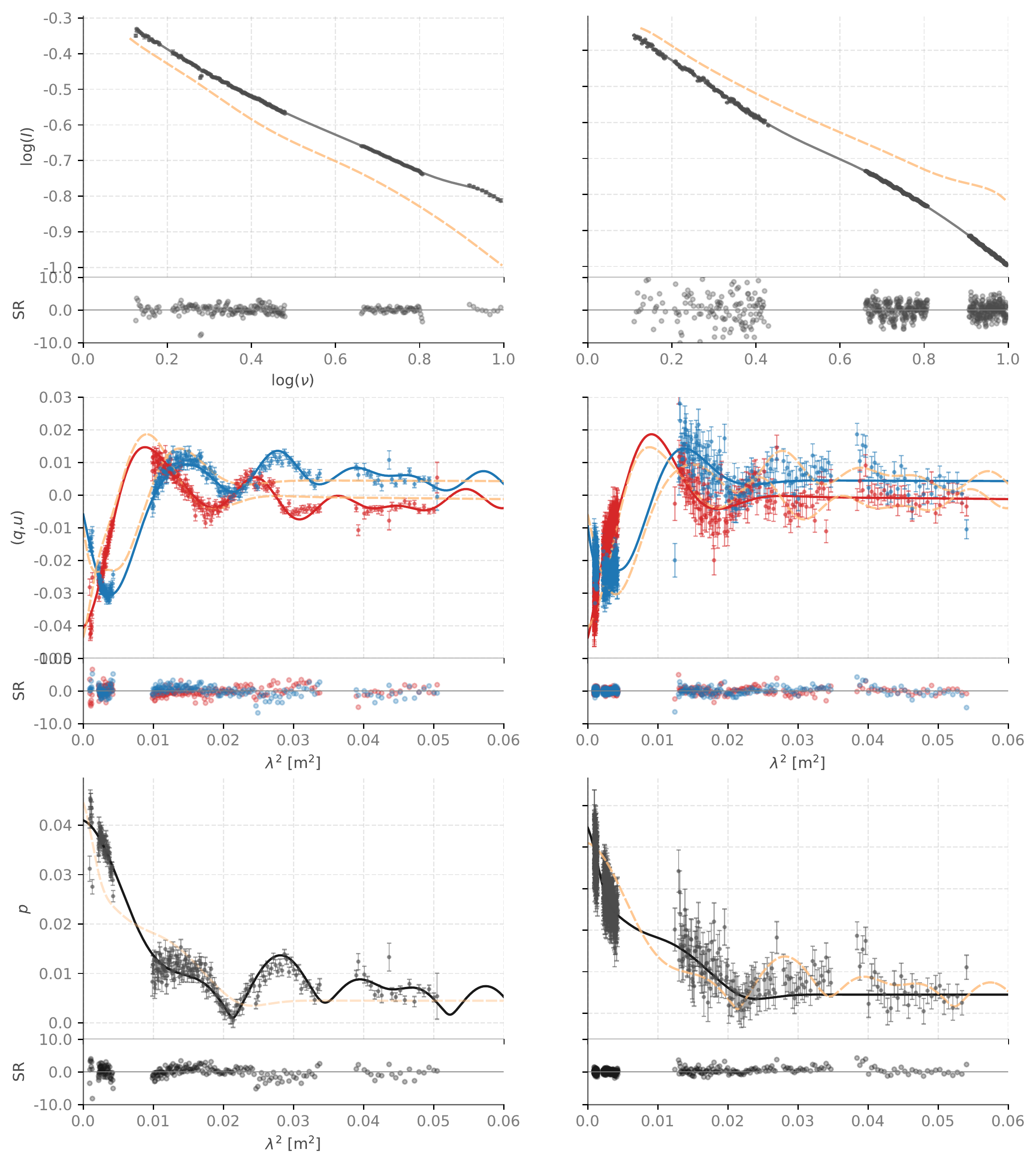}
\caption{As for Figure \ref{fig:0454_comparo}, but for PKS B0545-649.}
\label{fig:04_comparo}
\end{figure*}

\begin{figure*}[htpb]
\centering
\includegraphics[width=0.8\textwidth]{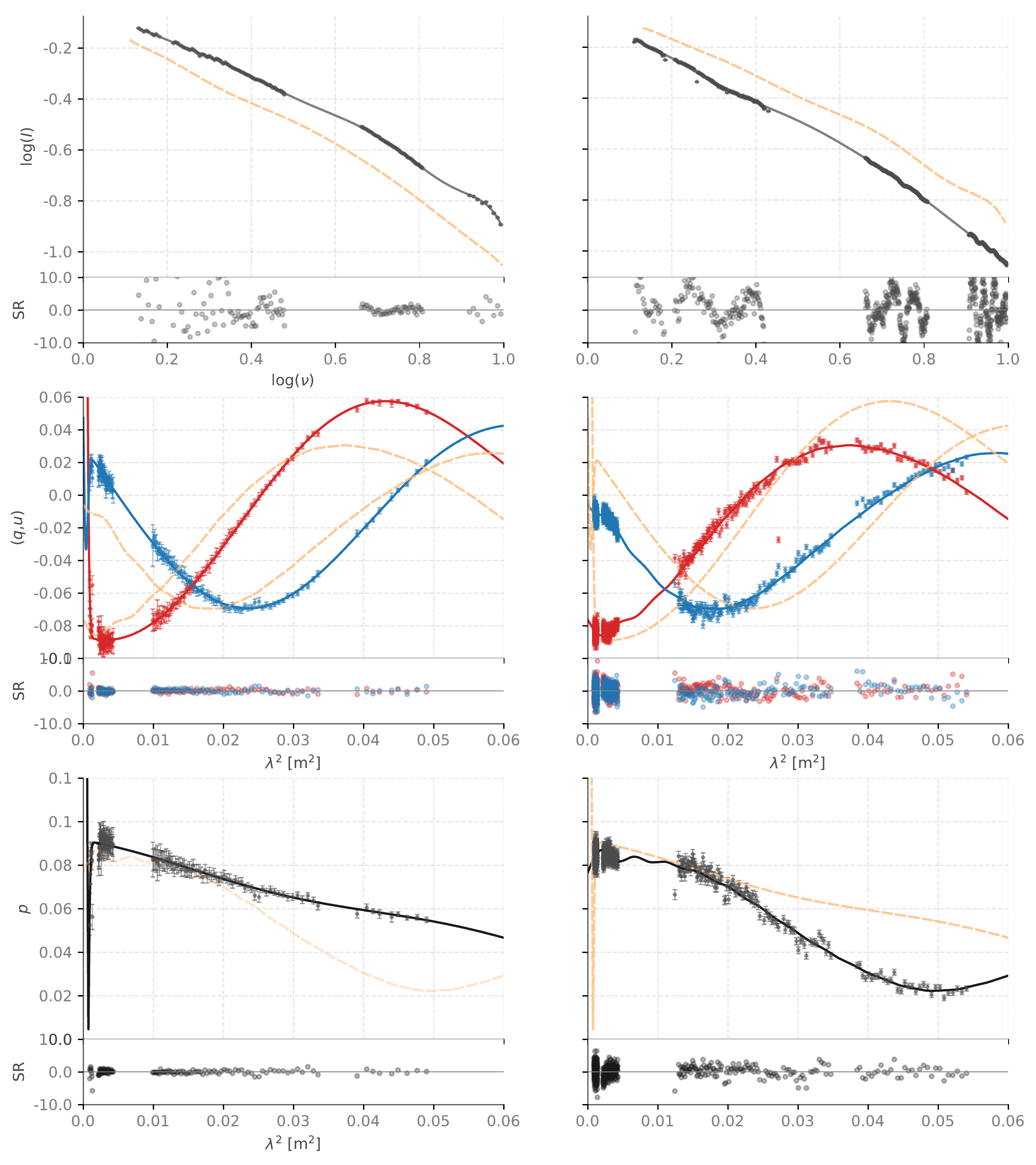}
\caption{As for Figure \ref{fig:0454_comparo}, but for PKS B0611-74}
\label{fig:03_comparo}
\end{figure*}

\begin{figure*}[htpb]
\centering
\includegraphics[width=0.8\textwidth]{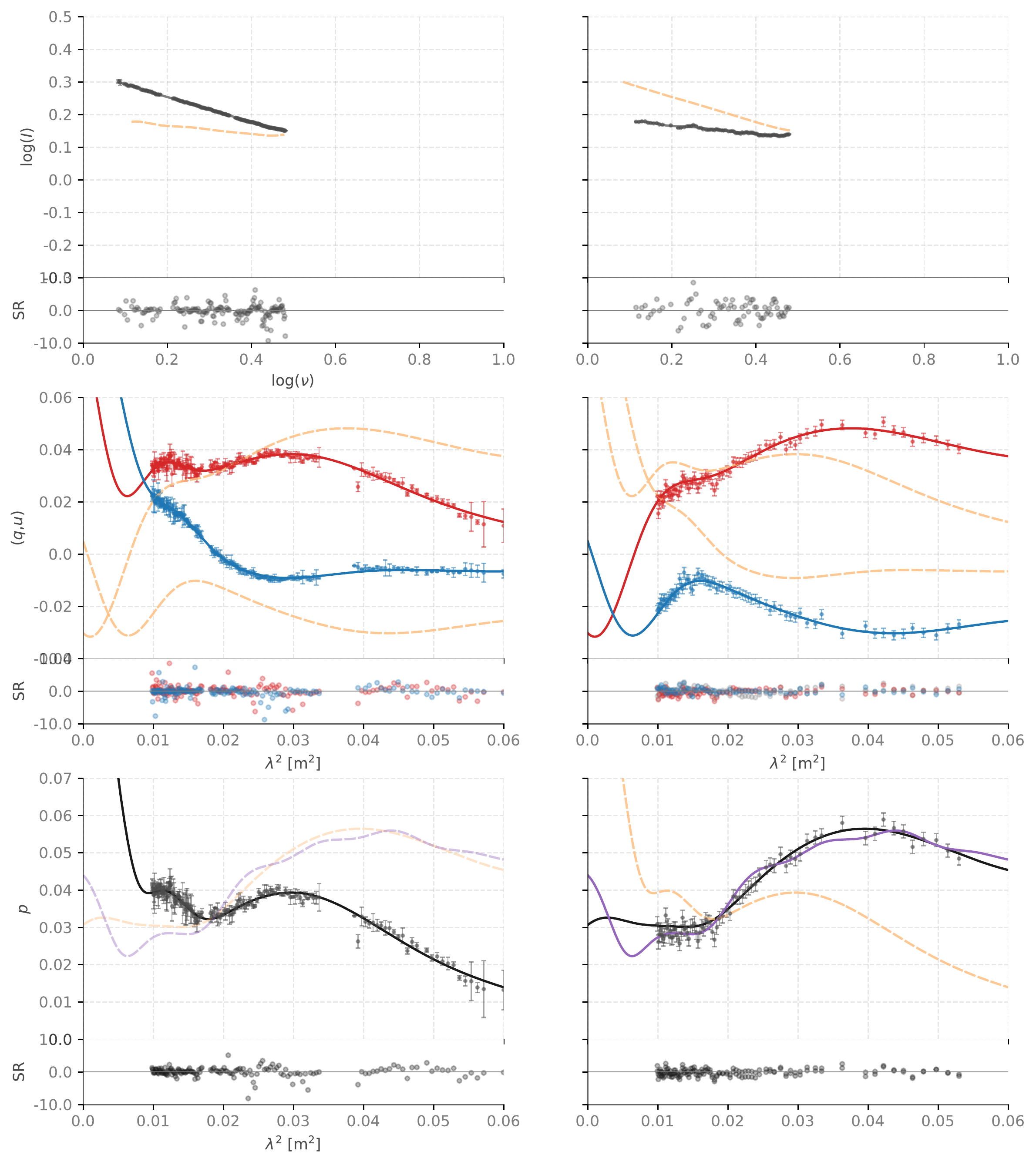}
\caption{As for Figure \ref{fig:0454_comparo}, but for PKS B1039-47}
\label{fig:1039_comparo}
\end{figure*}

\begin{figure*}[htpb]
\centering
\includegraphics[width=0.8\textwidth]{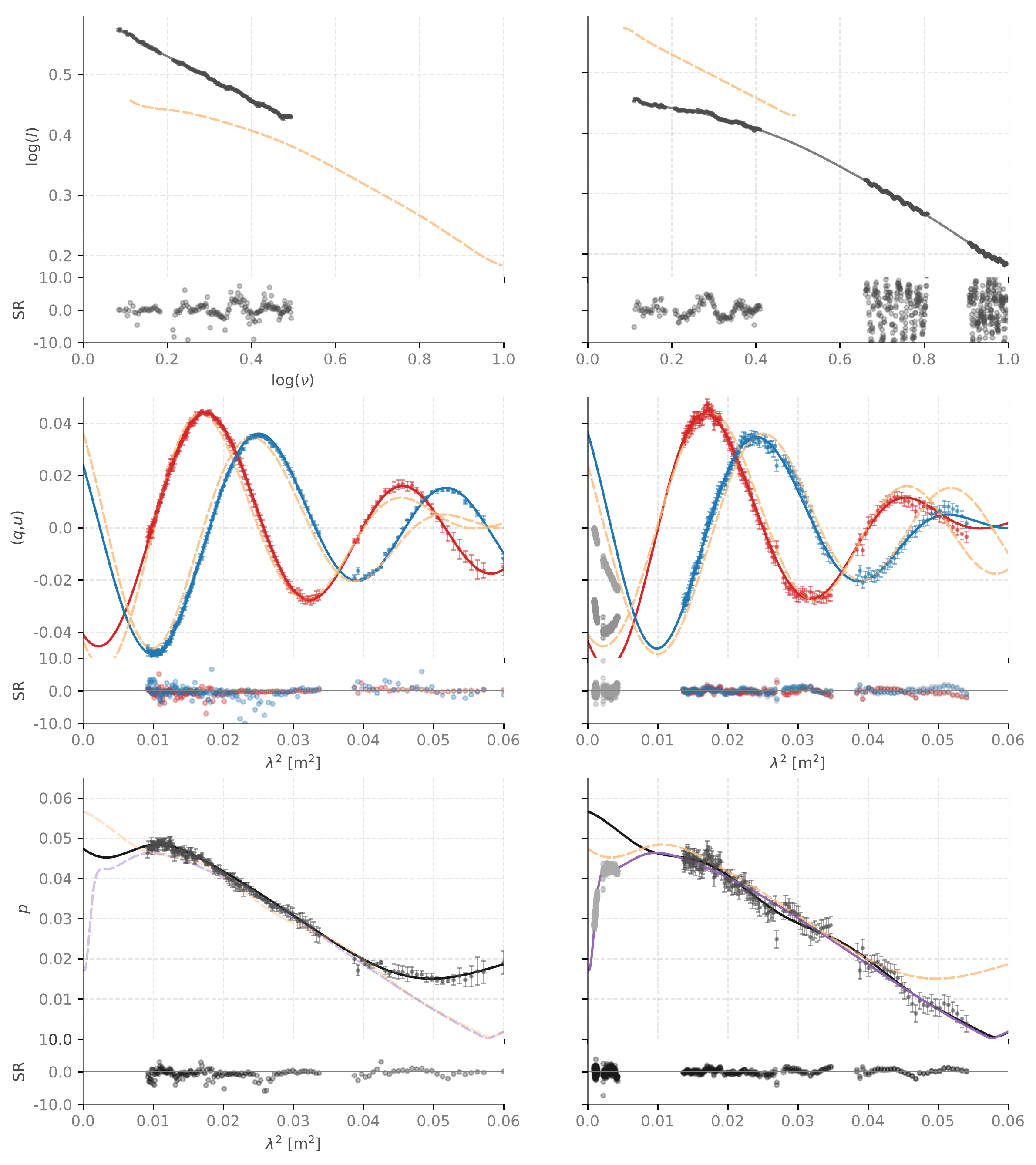}
\caption{As for Figure \ref{fig:0454_comparo}, but for PKS B1610-711}
\label{fig:1610_comparo}
\end{figure*}

\begin{figure*}[htpb]
\centering
\includegraphics[width=0.8\textwidth]{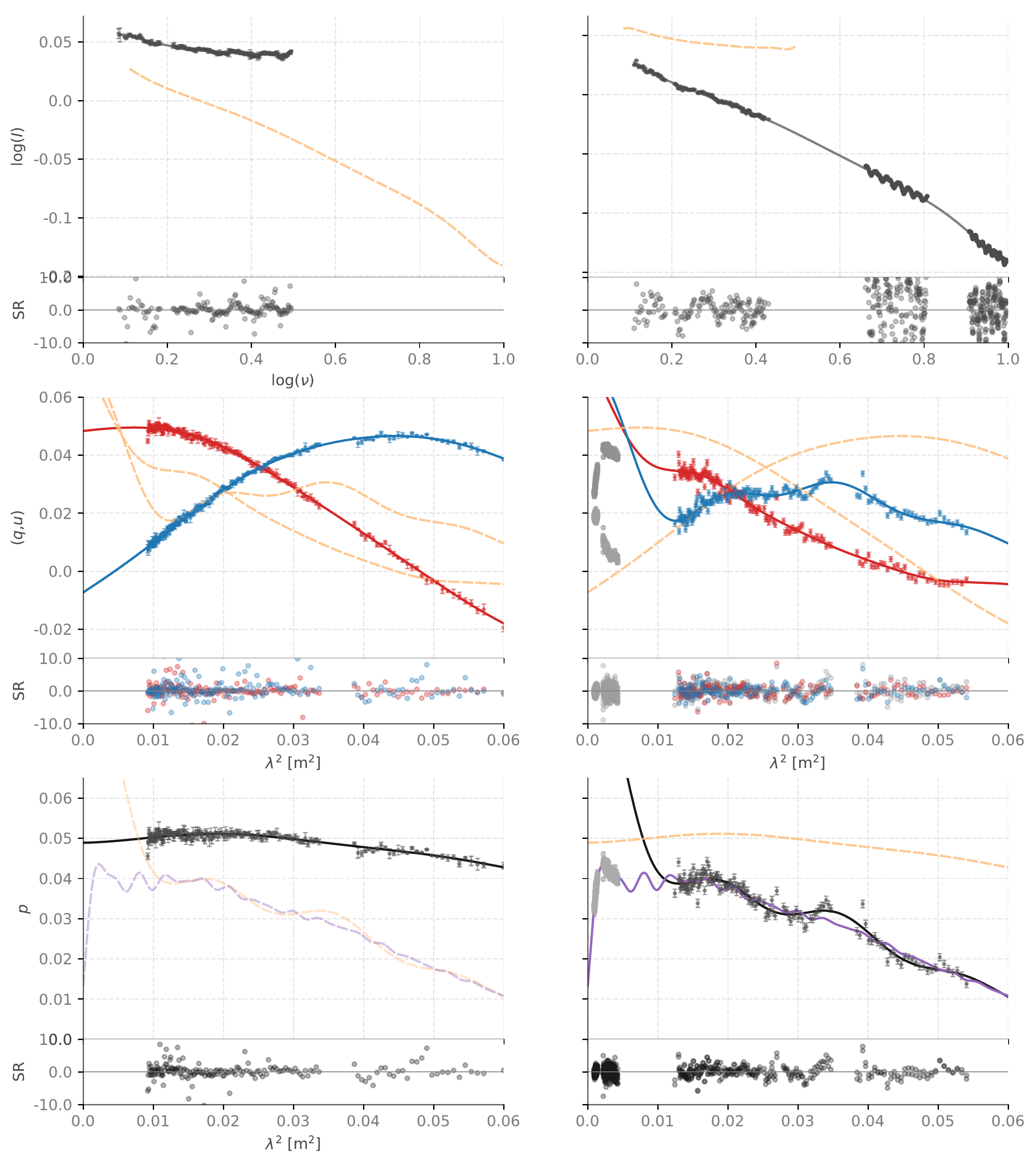}
\caption{As for Figure \ref{fig:0454_comparo}, but for PKS B1903-802.}
\label{fig:1903_comparo}
\end{figure*}

In Figures \ref{fig:0454_comparo}--\ref{fig:1903_comparo}, we present the Stokes $I$ spectrum (top row of axes), Stokes ($q$,$u$) spectrum (middle row), and fractional polarisation spectrum (bottom row) for each source in both the 2012 and 2017 epochs (left and right columns of axes respectively), along with our best-fit models to these data. The standardised residual (SR) between the data and the model fit, where SR $= (d-D_i)/\sigma_i$, and $d_i$, $D_i$, and $\sigma_i$ are the values of the data, is plotted on the associated sub-axes. For the OS12-selected sources, we have plotted the 2017 epoch 3/6cm band Stokes ($q$,$u$) and $p$ data in light gray to demonstrate the high frequency polarisation behaviour of the source, while making it clear that these data are excluded from the multi-epoch analysis. 

The standardised Stokes $I$ residuals show quasi-regular oscillations caused by the elevation difference between our bandpass calibrator and targets. This known issue with ATCA observations cannot easily be calibrated out, beyond simply dividing the oscillations out of the spectra (which we do not do). Nevertheless, the fractional amplitude of the oscillations is never larger than two per cent of Stokes $I$, so the same limit must apply to the Stokes $q$ and $u$ spectra. We verified that our results were unaffected by this aberration by re-running our analysis on Stokes $(q,u)$ data formed by dividing Stokes $Q$ and $U$ by $I$ on a channel-by-channel basis (cf Section \ref{sec-obs}). 

Our $(q,u)$ fitting results are recorded in Table \ref{table:BFModResults}. Columns 1--13 of the table contain (in order): The source name, the epoch of observation, the fitted Stokes $I$ spectral index and its uncertainty, the  Stokes $I$ flux density at 1.4 GHz (denoted $I_\text{1.4 GHz}$) to three decimal places, the best-fit model type, the base-10 logarithm of the Bayes factor, which is computed with respect to the model type with the next-highest marginal likelihood value, a list of other plausible candidate model types (see Section \ref{sec-comparison}), the best-fit parameter values and their uncertainties for $p_{0}$, $\psi_{0}$, RM, $\sigma_{\text{RM}}$, and $\Delta \phi$ for the $j$-th emission component in the best-fit model, and the reduced-$\chi^2$ (denoted $\tilde{\chi}^2$) of the best-fitting model. We note that the type and number of emission components in the best-fit model can be inferred on the basis of blanked entries in the table.  

\begin{turnpage}
\begin{deluxetable*}{lllllll} 
\tabletypesize{\footnotesize} 
\tablecolumns{8} 
\tablewidth{0pt} 
\tablecaption{Results from polarimetric modelling fitting and comparison (\emph{Table 3 continues on the following page})} 
\tablehead{ 
\colhead{(1)} & \colhead{(2)} & \colhead{(3)} & \colhead{(4)} & \colhead{(5)} & \colhead{(6)} & \colhead{(7)}\\ 
\colhead{Source} & \colhead{Epoch} & \colhead{$\alpha$} & \colhead{$I_\text{1.4 GHz}$} & \colhead{Best-fit} & \colhead{Logarithm of the Bayes} & \colhead{Plausible altenative} \\ 
\colhead{} & \colhead{} & \colhead{} & \colhead{} & \colhead{model type} & \colhead{Factor margin}  & \colhead{model type(s)} \\
\colhead{} & \colhead{} & \colhead{} & \colhead{[Jy]} & \colhead{}  & \colhead{}  & \colhead{}
}
\startdata 
B0454-810 & 2012 & 0.445 & 0.813 & IE & 0.3 & ME,MI,MIE \\ 
B0454-810 & 2017 & 0.349 & 0.971 & RRM & 0.6 & MII \\ 
B0454-810 & 2017* & - & - & III & 4.9 & -- \\ 
B0517-726 & 2012 & 0.316 & 0.202 & III & 13.4 & -- \\ 
B0517-726 & 2017 & -0.093 & 0.259 & RII & 487.6 & -- \\  
B0543-735 & 2012 & -0.432 & 0.994 & III & 12.5 & -- \\ 
B0543-735 & 2017 & -0.489 & 0.997 & III & 120.8 & -- \\ 
B0545-649 & 2012 & -0.552 & 0.448 & RRI & 1.8 & -- \\ 
B0545-649 & 2017 & -0.688 & 0.410 & REE & 48.6 & -- \\ 
B0611-74 & 2012 & -0.787 & 0.737 & IEE & 0.4 & IIE \\ 
B0611-74 & 2017 & -0.997 & 0.631 & RII & 10.2 & -- \\  
B1039-47 & 2012 & -0.380 & 1.883 & MEE & 1.4 & MMM \\ 
B1039-47 & 2017 & -0.127 & 1.499 & RMI & 0.8 & MII,MIE,MEE,EEE,RRM,RMM,IEE \\ 
B1039-47 & 2017* & - & - & III & 0.2 & RRI \\ 
B1610-771 & 2012 & -0.372 & 3.549 & RMI & 0.1 & RME,RIE \\ 
B1610-771 & 2017 & -0.158 & 2.800 & II & 0.1 & RMI,RI \\ 
B1610-771 & 2017* & - & - & MIE & 4.7 & -- \\ 
B1903-802 & 2012 & -0.041 & 1.126 & IE & 0.3 & ME,MI \\ 
B1903-802 & 2017 & -0.144 & 1.046 & MII & 0.6 & MMI \\ 
B1903-802 & 2017* & - & - & III & 5.5 & -- \\ 
\enddata 
\tablecomments{Both the spectral index and $I_\text{1.4 GHz}$ values have been quoted to 3 decimal places, which approximately reflect the limits placed on the variability of our flux calibrator PKS B1934-638 over a moving five year window by \citet{Tingay2003}. The formal fitting uncertainties are considerably lower. A `*' symbol against the epoch indicates that these results pertain to the full 1.3--10 GHz band, rather than the restricted 1.3.--3.1 GHz band (i.e. see Section \ref{sec-modelling}; also column 5 of Table \ref{tab:SourceDat}).} 
\label{table:BFModResults}
\end{deluxetable*}
\end{turnpage}

\addtocounter{table}{-1}
\begin{turnpage}
\begin{deluxetable*}{llllllll} 
\tabletypesize{\footnotesize} 
\tablecolumns{8} 
\tablewidth{0pt} 
\tablecaption{\emph{Table 3 continued} -- Results from polarimetric modelling fitting and comparison}
\tablehead{ 
\colhead{(1)} & \colhead{(2)} & \colhead{(8)} & \colhead{(9)} & \colhead{(10)} & \colhead{(11)} & \colhead{(12)} & \colhead{(13)}\\ 
\colhead{Source} & \colhead{Epoch} & \colhead{$p_{0,[j=1,2,3]}$} &  \colhead{$\psi_{0,[j=1,2,3]}$} & \colhead{RM$_{[j=1,2,3]}$} & \colhead{$\sigma_{\text{RM}[j=1,2,3]}$} & \colhead{$\Delta\phi_{[j=1,2,3]}$} & \colhead{$\tilde{\chi}^2$}\\ 
\colhead{}  & \colhead{} & \colhead{} & \colhead{} & \colhead{} & \colhead{} & \colhead{} & \colhead{}\\
\colhead{} & \colhead{}  & \colhead{} & \colhead{[rad]} & \colhead{[rad m$^{-2}$]} & \colhead{[rad m$^{-2}$]} & \colhead{[rad m$^{-2}$]} & \colhead{} 
}
\startdata 
B0454-810 & 2012 & 0.022(3),0.0389(8), -   & 0.4(1),-0.86(2),  -   & 60(10),38.4(6), -   &  64(6) ,   -   ,   -    &    -   , 7.1(6),   -   & 2.3 \\
B0454-810 & 2017 & 0.01(1),0.016(1),0.14(8) & -0.7(1),0.76(8), 0(1) & 6(4),56(2),400(100) &    -   ,   -   , 88(8)  &    -   ,   -   , 86(8)  & 1.0 \\
B0454-810 & 2017* & 0.01341(6),0.02057(8), 0.0108(1) & -0.894(4),0.73(3),-0.132(4) & -112(3),56.6(2),-11.3(3) &    -   ,   -   ,   -    &  365(2),  7(3) , 9.5(7) & 1.6 \\
B0517-726 & 2012 & 0.037(1),0.005(8),0.0446(9) & 0.82(2),0.6(7),-0.46(2) & 76.8(7),530(80),66.4(8) &    -   ,   -   ,   -    & 35.1(7),280(40),  3(2) & 1.4 \\
B0517-726 & 2017 & 0.0099(1),0.04078(4),0.02463(7) & -0.164(5),-1.413(3),0.326(6) & 69.7(4),77.2(4),49(3) &    -   ,   -   ,   -    &    -   ,71.1(4), 493(1) & 2.0 \\
B0543-735 & 2012 & 0.0087(2),0.079(1),0.129(1) & -1.16(2),-1.541(6),-0.125(4) & -498(7),-12(5),9.9(3) &    -   ,   -   ,   -    &  152(3),23.2(2),32.6(3) & 0.3 \\
B0543-735 & 2017 & 0.06(1),0.059(2),0.023(4) & 0.66(1),-0.374(5),-0.2(1) & -489(4),-414(3),45.7(5) &    -   ,   -   ,   -    &  308(3), 470(3),14.3(3) & 2.3 \\
B0545-649 & 2012 & 0.005(1),0.0021(9),0.0352(2) & 1.33(2),1.44(4),-1.445(4) & -9.8(9),316(1),133.7(5) &    -   ,   -   ,   -    &    -   ,   -   ,55.2(3) & 2.3 \\
B0545-649 & 2017 & 0.00446(8),0.0179(4),0.0313(2) & 0.78(2),-1.17(1),-1.553(4) & 2.4(8),277(9),153.1(9) &    -   , 267(6),52.6(5) &    -   ,   -   ,   -   & 1.6 \\
B0611-74 & 2012 & 0.66(3),0.082(4),0.033(4) & 0(2),-1.55(3),0.84(7) & -200(200),34.7(6),26(7) & 1340(10), 9.9(6), -  &   - , - ,24(1) & 0.4 \\
B0611-74 & 2017 & 0.0392(4),0.0076(1),0.077(1) & 0.92(2),-0.65(1),-1.326(2) & 53.5(5),-400(5),35.7(3) &  - , - , -  &  - ,175(3),28.(2) & 6.1 \\
B1039-47 & 2012 & 0.0511(8),0.019(1),0.06(1) & 0.2(2),0.64(5),0.5(2) & -7.5(4),60(2),110(20) & 13.3(1), 23(1) , 56(6)  &    -   ,   -   , 94(4) & 1.7 \\
B1039-47 & 2017 & 0.0481(4),0.039(2),0.04(1) & -0.17(2),1.47(7),1.4(3) & -2.4(6),-7(3),110(30) &    -   ,   -   , 61(9)  &    -   ,36.1(5), 30(20) &  1.1 \\
B1039-47 & 2017* & 0.0492(2),0.026(2),0.0306(5) & -0.162(9),0.1(1),-1.51(2) & -2.2(3),62(8),-14.3(9) &    -   ,   -   ,   -    &  0.7(4), 147(2),33.7(4) & 1.2  \\
B1610-771 & 2012 & 0.018(2),0.035(2),0.008(3) & 1.36(8),1.37(4),0.2(3) & 113(1),90(2),-0(30) &    -   ,   -   , 60(20) &    -   , 17(2) , 40(20) & 1.0 \\
B1610-771 & 2017 & 0.013(2),0.0487(5), -   & 0.8(2),1.32(1),  -   & 72(7),104.6(5), -   &    -   ,   -   ,   -    &  90(2) ,17.2(1),   -   & 0.5 \\
B1610-771 & 2017* & 0.0151(6),0.0495(3),0.0316(6) & -0.79(2),1.368(7),0.09(2) & 163(7),103(3),-0(20) &  123(3),   -   ,600(10) &    -   ,17.4(1),130(70) & 0.8 \\
B1903-802 & 2012 & 0.0523(2),0.0035(3), -  & -0.066(4),-1.5(1),  -   & 17.8(1),38(7), -   &  5.3(1),   -   ,   -    &    -   , 50(3) ,   -  & 1.3 \\
B1903-802 & 2017 & 0.0426(2),0.025(2),0.05(6) & 0.079(6),0.3(1),0.69(8) & 15.8(2),-25(3),40(4) &    -   ,   -   , 23(3)  & 13.2(8),97.8(5),93.8(5) & 3.4 \\
B1903-802 & 2017* & 0.04138(8),0.0131(4),0.022(4) & 0.003(1),1.12(2),-1.49(1) & 17.99(8),-572(3),-270(2) &    -   ,   -   ,   -    & 13.02(6), 260(2), 272(1) & 2.3
\enddata 
\tablecomments{A `*' symbol against the epoch indicates that these results pertain to the full 1.3--10 GHz band, rather than the restricted 1.3.--3.1 GHz band (i.e. see Section \ref{sec-modelling}; also column 5 of Table \ref{tab:SourceDat}).} 
\label{table:BFModResults}
\end{deluxetable*}
\clearpage
\end{turnpage}

 A summary of our basic fitting results is as follows:

\begin{itemize}
\item We obtain reasonable fits to the ($q$,$u$) data, with minimum, mean, and maximum $\tilde{\chi}^2$ values of 0.4, 1.7, and 6.1 respectively. Most of the $\tilde{\chi}^2$ values are slightly elevated above 1.
\item The best-fit model types for our sources often contain three emission components. 
\item The last two points may indicate that more emission components are needed to fully describe the polarised signal for our sample. Conversely, we note that the standardised residuals typically do not show obvious frequency-dependant structure, and are small in magnitude over the full band. 
\item The different types of emission components (i.e. R, I, E, and M) are all represented in the best-fit model types, and somewhat evenly.
\item For the fitted model components, we derive median values for $p_0$, $|$RM$|$, $\sigma_\text{RM}$, and $\Delta\phi$ of 3.1 percent, 56 rad m$^{-2}$, 59 rad m$^{-2}$, and 43 rad m$^{-2}$ respectively, and ranges of (again, respectively) 0.2--66 per cent, 0.4--570 rad m$^{-2}$, 5--1350 rad m$^{-2}$, and 0--495 rad m$^{-2}$ (in the last two cases, where components of this type were present in the best-fit model, and remembering the bandwidth differences between the OS12- and A16-selected sources).
\item The frequency-dependent polarisation structure of several sources differs between epochs, as do their best-fit model types and parameters. 
\end{itemize}

\noindent We examine this last claim more rigorously in the next section, then describe the nature of the observed changes in Section \ref{sec-bfmodelcomparo}.

\subsection{Discounting changes induced by calibration and observational uncertainties}\label{sec-bfmodelcomparo} 

It is evident that there are differences in the polarisation spectra between epochs, but these could be due to calibration uncertainties or observational effects. To assess this possibility, we took the best-fit model for each source in the 2012 epoch, allowed some of its defining parameter values to vary by an amount commensurate with relevant calibration or observational uncertainties, then determined whether this could provide a good fit to the 2017 epoch data. Specifically, we took the equation defining the best-fit model for the 2012 epoch, then made the substitutions:

\begin{equation*}
\psi_{0,[j]} \rightarrow \psi_{0,[j],\text{bf}}+\delta \psi_0
\end{equation*}

\begin{equation*}
\text{RM}_{[j]} \rightarrow \text{RM}_{[j],\text{bf}}+\delta \text{RM}
\end{equation*}

\noindent where the `bf' subscript indicates parameter values that yield the best-fit model for the 2012 epoch. We then fit for the `$\delta$' parameters, subject here to informative priors, using the same methods described in Section \ref{sec-fitting}. We adopted Gaussian priors for $\delta \psi_0$ and $\delta \text{RM}$, with full-width-half-maximums of five degrees and 10 rad m$^{-2}$ respectively --- more than accounting for the uncertainty in the online angle calibration of ATCA (accurate to one degree), and typical variation in ionospheric Faraday rotation (of order a few rad m$^{-2}$ --- e.g. \citealp{SotBel2013,Lenc2017}). Since all emission components are equally affected by errors in the angle calibration and ionospheric Faraday rotation, $\delta\psi_0$ and $\delta \text{RM}$ are forced to be identical for all $j$ components in the adjusted 2012 model. We do not attempt to fit for calibration uncertainties on the $p_{0,[j]}$ parameters: Our absolute flux scale is uncertain by $\sim10$\% (see Section \ref{sec-obs}), but Stokes $I$, $Q$ and $U$ are all tied to this scale, so calibration uncertainties in our fractional polarisation values are dominated by the (small) leakage values reported in Section \ref{sec-obs}. Neither do we fit for calibration uncertainties on the $\sigma_\text{RM}$ or $\Delta \phi$ parameters, but in this case, because we cannot think of instrumental or observational issues that would plausibly mimic changes in these quantities between epochs. 

We find that incorporating calibration uncertainties into the 2012 best-fit models provides a plausible fit to the 2017 data for PKS B0515-674 only --- an essentially unpolarised source. All other sources show statistically significant change between epochs.

\subsection{Variability results and notes for individual sources}\label{sec-bfmodelcomparo}

\begin{figure*}[htpb]
\centering
\includegraphics[width=\textwidth]{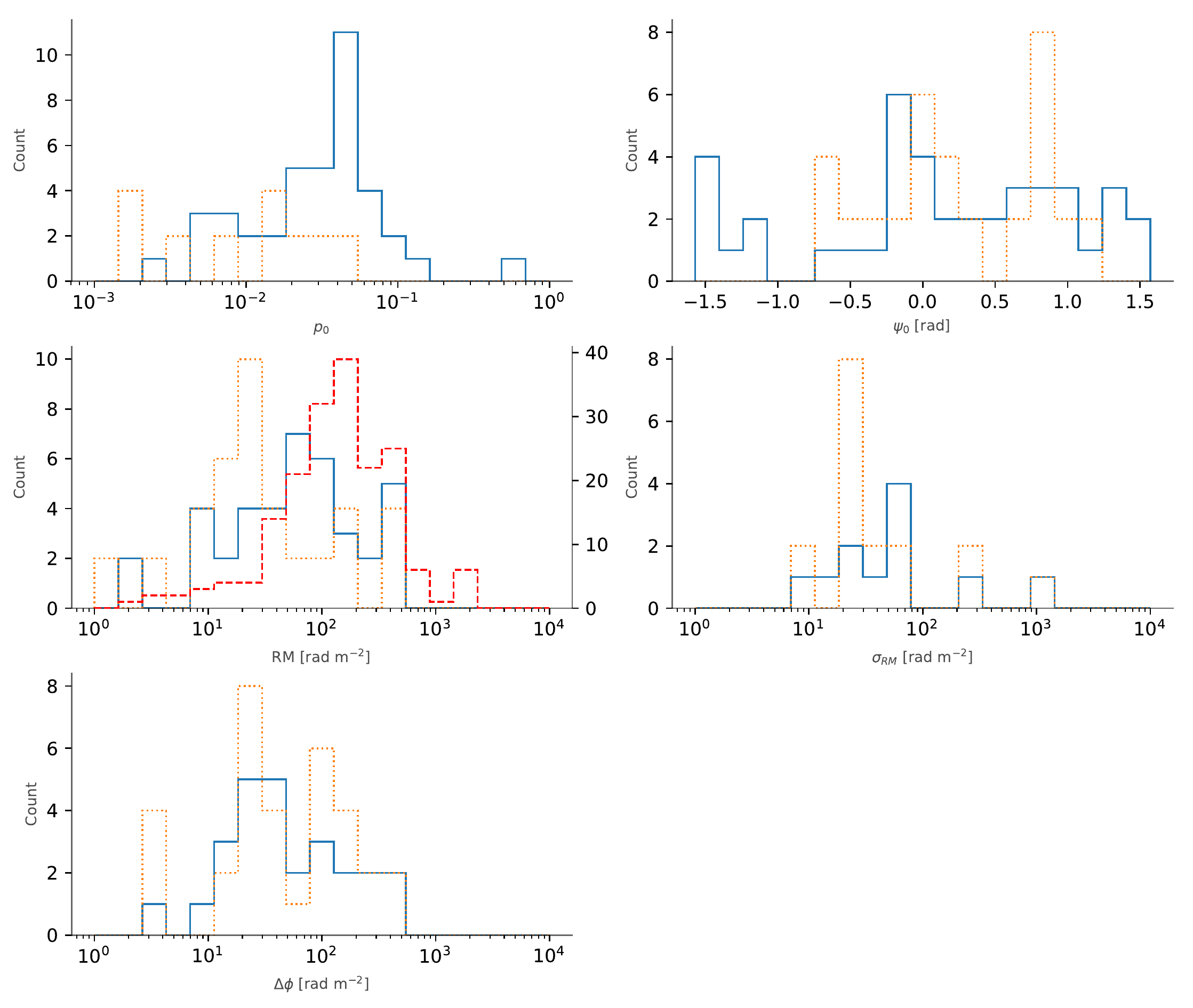}
\caption{Histograms of the best-fit values for $p_0$, $|$RM$|$, $\sigma_\text{RM}$, and $\Delta\phi$ (blue lines, left-to-right, top-to-bottom) in both 2012 and 2017, and the change therein between epochs (orange dotted lines) for our cross-epoch component associations (see Section \ref{sec-bfmodelcomparo} and Table \ref{table:compAssociationResults}). The red dashed histogram on the third axis shows the distribution of $|$RM$|$ in isolated jet components in the \citet{Hovatta2012} blazar sample, located up to 100 pc from the central SMBH (see Figure 4 of that work).}
\label{fig:pvdplots}
\end{figure*}

In this section we discuss our fitting results for sources on a case-by-case basis. We highlight results from the literature that are relevant for constraining source structure, or for establishing prior evidence for polarimetric variability. Also, since there is no \emph{a priori} reason to expect that a one-to-one mapping will consistently occur between our $j$-indexed model components (i.e. from Eqn. \ref{eqn:ShaneMod} and Table \ref{table:BFModResults}) and specific physical emission regions across epochs, we attempt to make such associations manually in this section. We do so by assuming that polarisation characteristics of the emitting regions have undergone the minimum level of observable change, then manually searching for similarities in the best-fit parameter values of the different $j$-indexed model components across epochs. Where the best-fit values for one or more model parameters across epochs are similar, we claim that these model components are describing emission from the same physical region. These associations must be regarded as tentative, but are useful as a foil for discussion. We provide our reasoning for each source. The results of these associations are listed in Table \ref{table:compAssociationResults}: In columns 1--12, we record (respectively) a designation for each component association (given in sections \ref{sec-indivFirstSource}--\ref{sec-indivLastSource}), the host source designation, the value of $p_0$ in the 2012 epoch and the change therein between 2012 and 2017, the value of $\psi_0$ in 2012 and its change between 2012 and 2017, and likewise for the RM, $\sigma_{\text{RM}}$, and $\Delta\phi$ parameters. We illustrate these results visually in Figure \ref{fig:pvdplots}, where both the best-fit values of the parameters and their change between epochs are plotted as histograms. Finally, we also provide an example of how the changes in the best-fit model of each source might be interpreted as a physical change in the source itself. These are intended to be illustrative only, are not unique, and are provided merely as examples to aid the reader.

\begin{turnpage}
\begin{deluxetable*}{llllllllllll}
\tabletypesize{\scriptsize} 
\tablecolumns{12} 
\tablewidth{0pt} 
\tablecaption{Change in associated components across epochs}
\tablehead{ 
\colhead{(1)} & \colhead{(2)} & \colhead{(3)} & \colhead{(4)} & \colhead{(5)} & \colhead{(6)} & \colhead{(7)} & \colhead{(8)} & \colhead{(9)} & \colhead{(10)} & \colhead{(11)} & \colhead{(12)}\\ 
\colhead{Component} & \colhead{Host source} & \colhead{$p_0$ (2012)} & \colhead{$\Delta p_0$}  & \colhead{$\psi_0$ (2012)} & \colhead{$\Delta \psi_0$}  & \colhead{RM (2012)} & \colhead{$\Delta$ RM}  & \colhead{$\sigma_{\text{RM}}$ (2012)} & \colhead{$\Delta ~\sigma_{\text{RM}}$}  & \colhead{$\Delta\phi$ (2012)} & \colhead{$\Delta ~\Delta\phi$}\\
\colhead{} & \colhead{} & \colhead{} & \colhead{}  & \colhead{[rad]} & \colhead{[rad]}  & \colhead{[rad m$^{-2}$]} & \colhead{[rad m$^{-2}$]}  & \colhead{[rad m$^{-2}$]} & \colhead{[rad m$^{-2}$]}  & \colhead{[rad m$^{-2}$]} & \colhead{[rad m$^{-2}$]}
}
\startdata 
 C1  & B0517-726 & 0.037(2)       & 0.004(2)   & 0.82(3)        & -2.23(3)  & 76.8(8)        & 0.4(9)      & $\bullet\quad$ & $\bullet\quad$ & 35(1)          & 36(1)          \\
 C2  & B0517-726 & 0.005(9)       & -0.035(2)  & 0.6(9)         & 0.30(3)   & 5.3(9)e+02     & 3(1)        & $\bullet\quad$ & $\bullet\quad$ & 2.8(4)e+02     & -3(2)          \\
 C3  & B0517-726 & 0.045(2)       & 0.020(9)   & -0.46(3)       & -0.3(9)   & 66.4(9)        & -4.8(9)e+02 & $\bullet\quad$ & $\bullet\quad$ & 3(2)           & 2.1(4)e+02     \\
 C4  & B0543-735 & 0.0087(2)      & 0.052(6)   & -1.16(2)       & 1.83(4)   & -498(7)        & 1(1)e+01    & $\bullet\quad$ & $\bullet\quad$ & 152(3)         & 1.6(1)e+02     \\
 C5  & B0543-735 & 0.079(1)       & -0.02(5)   & -1.541(6)      & 1.167(8)  & -12(5)         & -402(6)     & $\bullet\quad$ & $\bullet\quad$ & 23.2(2)        & 447(3)         \\
 C6  & B0543-735 & 0.129(1)       & -0.106(1)  & -0.125(4)      & -0.1(5)   & 9.9(3)         & 36(2)       & $\bullet\quad$ & $\bullet\quad$ & 32.6(3)        & -18.2(5)       \\
 C7  & B0545-649 & 0.005(1)       & -0.001(1)  & 1.33(2)        & -0.55(4)  & -9.8(9)        & 12(1)       & $\bullet\quad$ & $\bullet\quad$ & $\bullet\quad$ & $\bullet\quad$ \\
 C8  & B0545-649 & 0.0021(1)      & 0.0158(4)  & 1.44(4)        & -2.61(4)  & 316(2)         & -39(9)      & $\bullet\quad$ & 267(7)         & $\bullet\quad$ & $\bullet\quad$ \\
 C9  & B0545-649 & 0.0352(2)      & -0.0039(3) & -1.445(4)      & -0.108(6) & 133.7(5)       & 19(1)       & $\bullet\quad$ & 52.6(5)        & 55.2(3)        & -55.2(3)       \\
 C10 & B0611-74  & 0.66(5)        & -0.005(5)  & 0(2)           & 1(2)      & -2(2)e+02      & 1.0(8)      & 1.34(2)e+03    & -9.9(7)        & $\bullet\quad$ & 28.7(2)        \\
 C11 & B0611-74  & 0.082(5)       & 0.006(5)   & -2(2)          & 0.08(7)   & 34.7(7)        & 27(9)       & 9.9(7)         & $\bullet\quad$ & $\bullet\quad$ & -24(1)         \\
 C12 & B0611-74  & 0.033(5)       & -0.65(5)   & 0.84(7)        & -1(2)     & 26(9)          & -2(2)e+02   & $\bullet\quad$ & -1.34(2)e+03   & 24(1)          & 176(5)         \\
 C13 & B1039-47  & 0.05(2)        & -0.01(2)   & 0.2(2)         & 1.3(2)    & -1(3)e+01      & 0(3)e+01    & 14(5)          & -14(5)         & $\bullet\quad$ & 36.1(5)        \\
 C14 & B1039-47  & 0.02(2)        & -0.02(1)   & 0.5(2)         & 0(1)      & 6(3)e+01       & 1(4)e+01    & 21(5)          & 0(1)e+01       & $\bullet\quad$ & -6(2)e+01      \\
 C15 & B1039-47  & 0.06(1)        & 0.03(2)    & 0.5(2)         & -0.7(2)   & 1.1(2)e+02     & -6(3)e+01   & 56(6)          & -21(5)         & 94(5)          & $\bullet\quad$ \\
 C16 & B1610-771 & 0.018(3)       & 0.01(2)    & 1(1)           & -0.1(3)   & 112(3)         & 1(2)e+01    & $\bullet\quad$ & $\bullet\quad$ & $\bullet\quad$ & 0(4)e+01       \\
 C17 & B1903-802 & 0.01(2)        & 0.03(2)    & -0.1(7)        & 0.2(7)    & 2(2)e+01       & -0(2)e+01   & 3(2)e+01       & -3(2)e+01      & $\bullet\quad$ & 13.2(8)        \\
 C18 & B1903-802 & 0.05(2)        & -0.00(2)   & -0.1(7)        & 0.8(7)    & 2(1)e+01       & 2(1)e+01    & $\bullet\quad$ & 24(3)          & 1(2)e+01       & 8(2)e+01  
\enddata 
\tablecomments{Results for two-epoch component associations described in Sections \ref{sec-indivFirstSource}--\ref{sec-indivLastSource}. Columns 1 and 2 contain an assigned component designation (see Sections \ref{sec-indivFirstSource} -- \ref{sec-indivLastSource}) and the associated source respectively. For each component, the odd numbered columns (starting from column 3) contain the best-fit parameter values and their statistical uncertainties (in standard parentheses notation) for the intrinsic fractional polarisation, intrinsic polarisation angle, RM, foreground RM dispersion, and intrinsic width in Faraday depth. The adjacent even-numbered columns contain the change in these parameter values between the 2012 and 2017 epochs, assuming our component associations. Bullets indicate that a given parameter is not present in the 2012 and/or 2017 epoch best fit model types, depending on the columns in which they are found.} 
\label{table:compAssociationResults}
\end{deluxetable*}
\clearpage
\end{turnpage}

\subsubsection{PKS B0454-810}\label{sec-indivFirstSource}

This source is an FSRQ, resolved on scales of $\sim5$ mas \citep{Ojha2005,Dodson2008}, and located at $z=0.44$ (OS12). \citet{Tingay2003} observed it at 8.6, 4.8, 2.5, and 1.4 GHz in numerous epochs between October 1996 and February 2000, and reported that Stokes $I$ varied by 25\%, 21\%, 15\%, and 7\% (respectively) root-mean-squared (RMS) over the period, while the fractional polarisation varied from 0--6\% with a complex frequency dependence (see their Figure 2). Multi-epoch measurements of its fractional polarisation at higher frequencies (2.6(5)\% at 18.5 GHz in 2002 by \citealt{Ricci2004}; $<1.5$\% at 20 GHz in 2007 by \citealt{Murphy2010}) further establish a history of spectral complexity and temporal variability.

We achieve a good fit to the data with an IE-type model in 2012 and an RRM model in 2017 (Table \ref{table:BFModResults}), though there are plausible alternatives in both epochs. The well-constrained parameter values in the best-fit models have fractional polarisations of up to $\sim$a few per cent, RMs up to 400 rad m$^{-2}$, and Faraday dispersions of up to 90 rad m$^{-2}$. There are no obvious cross-epoch component associations for this source --- all parameter values change significantly between epochs --- so we do not attempt to make any. The IE to RRM model transition could correspond to, for example, the emergence of a new weak component with high RM and Faraday dispersion (the `M' component), and the propagation of the I and E -type components from behind complex parts of a foreground Faraday screen (where an RM gradient and turbulent RM foreground is imposed on them) to less structured parts of the foreground screen, which only induce Faraday rotation.

\subsubsection{PKS B0515-674} 

This source has not been well-studied. It is unresolved at 0.15" \citep{Chhetri2013} but has a steep spectrum ($\alpha=-0.76$ at 1.4 GHz; A16). It is effectively unpolarised in both epochs over 1.3--9.9 GHz, and remains undetected in polarisation up to 20 GHz \citep{Murphy2010}. Its Stokes $I$ spectrum remains unchanged over 1--3 GHz between 2012 and 2017, but has faded and steepened somewhat between 4 and 9.9 GHz. A detailed analysis of this behaviour is beyond the scope of this work, but suggests that a faint, GHz-peaked emission component has faded in the core of the AGN between epochs. Again, we make no cross-epoch associations for components in this source, due to its lack of polarised emission.

\subsubsection{PKS B0517-726}\label{sec:remarkable}

This source is an FSRQ \citep{Healey2007} sitting at an unknown redshift, and is spatially unresolved down to 0.15" \citep{Chhetri2013}. It was observed at 20, 8, and 5 GHz between October and November 2005 by \citet{Murphy2010}, and was found to have fractional polarisations of 2.4\%, 4.3\% and 4.0\% (respectively). In contrast, A16 report that the fractional polarisation was less than $2$\% over 4--10 GHz in 2012. Moreover, comparison of narrowband 1.4 GHz data from 1995 \citep{Gaensler2005} and 2012 A16 revealed an approximate doubling of the fractional polarisation and peak Faraday depth over the intervening period (see Section 6.3 of A16). The 2017 observations presented here reveal that the polarisation spectrum has changed once again, differing markedly from the 2012 epoch (Figure \ref{fig:13_comparo}), but essentially reverting to its 1995 state over the relevant frequency range. 

Table \ref{table:BFModResults} shows that the source is well-fit by an III-type model in 2012, and reasonably well-fit by an RII-type model in 2017. We associate the $j=1,3$ components from 2012 with the $j=2,1$ components (respectively) from 2017 to form components C1 and C2 in Table \ref{table:compAssociationResults} and Figure \ref{fig:pvdplots} respectively. This results in changes in the RM of both components, and in $\Delta\phi$ for C2, of less than 1\%. It leaves the $j=2$ and $j=3$ components from 2012 and 2017 (respectively) free to be associated as component C3. The main changes between epochs are the fractional polarisation of C2 (-3.5\%), the intrinsic angle of C1 (-2.2 rad), the RM of component C3 (-480 rad m$^{-2}$), and in the magnitude of $\Delta\phi$ for components C1 and C3 (+36 rad m$^{-2}$ and +210 rad m$^{-2}$). It is remarkable that the extracted RM values of components C1 and C2 differ so little across epochs, despite manifest differences in the raw polarisation spectra between epochs (Fig. \ref{fig:13_comparo}). This represents strong evidence that intrinsic source properties can be accurately extracted from high-quality 1-dimensional spectropolarimetric data. The III to RII model transition might correspond to, for example, an emission component with an RM gradient imposed by a foreground Faraday screen propagating to behind a simpler part of that screen, which induces Faraday rotation only.
 
\subsubsection{PKS B0543-735} 

This is a candidate $\gamma$-ray-emitting blazar \citep{DAbrusco2014}, but is mildly extended on 0.15" scales \citep{Chhetri2013}. \citet{Murphy2010} report fractional polarisations of 7.8\% , 6.2\% and 5.9\% at 20, 8, and 5 GHz between October and November 2005. In contrast, A16 measure 5.0\% and 5.2\% (averaged over 8--10 GHz and 4--6 GHz, respectively) in 2012. Comparison of narrowband 1.4 GHz polarisation data between 1995 and 2012 (see A16) revealed statistically-significant changes, including a halving of the fractional polarisation, a near-tripling of the peak Faraday depth, and a $\sim1$ rad change in the intrinsic polarisation angle. 

The source is well-fit by an III model in 2012, and reasonably so by the same model type in 2017, though in the latter case we note the existence of clear structure in the residuals for $(q,u)$. This suggests that an additional emission component is required to fully describe its behaviour. Nevertheless, we associate the $j=1,2,3$ components with themselves across epochs to form components C4, C5, and C6 in Table \ref{table:compAssociationResults} and Figure \ref{fig:pvdplots}, which yields a small fractional change in RM for C4 (2\%), in $p_0*I$ for C5 (-2\%) and in the values of $\psi_0$ and $\Delta\phi$ for C6 ($<0.1$ rad and -18 rad m$^{-2}$, respectively). As a result, the main changes are in the fractional polarisation of component C4 (+5.2\%), the intrinsic angles of components C4 \& C5 (-1.3 rad, +1.17 rad), the RMs of components C5 \& C6 (-400 rad m$^{-2}$, +36 rad m$^{-2}$), and in the values of $\Delta\phi$ for components C4 \& C5 (+160 rad m$^{-2}$ and +447 rad m$^{-2}$). The I-type emission components in the best-fit models might be associated with RM gradients imposed on the components by a foreground Faraday screen, or components in which the synchrotron-emitting and Faraday-rotating plasmas are mixed together.

\subsubsection{PKS B0545-649}\label{sec-borderline}

\citet{Murphy2010} measured the fractional polarisation of this source to be 2.7\% at 5 GHz (and supply upper limits of 3.5\% and 4.9\% at 8 and 20 GHz respectively). A16 measure a marginally higher value in 2012 of $\sim3$\% at 5 GHz. This might be due to a consistent fading in Stokes $I$ of $\sim6$ mJy yr$^{-1}$ (\citealt{Murphy2010} measure log($I)=-0.62$ at 5 GHz in 2006; we measure log($I)=-0.68$ in 2012 at the same frequency, and log($I)=-0.76$ at 5 GHz in 2017).

The source is well-fit by an RRI model in 2012 and an REE model in 2017, with no plausible alternatives in either epoch. We again associate the $j=1,2,3$ components with themselves across epochs to form components C7, C8, and C9 in Table \ref{table:compAssociationResults} and Figure \ref{fig:pvdplots}. This yields a small change in the intrinsic polarisation angle of C9 (-0.11 rad), and relatively small changes in the RM of components C7 \& C9  (+12 rad m$^{-2}$, and +19 rad m$^{-2}$ corresponding to a 15\% change in the latter case). We note that the $\Delta\phi$ value of component C9 decreases by 55 rad m$^{-2}$ but a nearly identical amount is gained in its $\sigma_\text{RM}$ value in the 2017 epoch --- this perhaps reflects a degeneracy in the modelling, and not a true change in the source. The main changes in the source are therefore a tenfold increase in the fractional polarisation of C8, a $-0.55$ rad and +0.64 rad change in $\psi_0$ for C7 and C8, a $-40$ rad m$^{-2}$ change in RM for C8, and a +270 rad m$^{-2}$ change in the $\sigma_\text{RM}$ value of C8. The RRI to REE model transition might result from two of the emission components transiting behind a more complicated part of a foreground Faraday screen. That is, two components transition to experiencing Faraday rotation by a complex turbulent foreground medium --- one having previously experienced only uniform foreground Faraday rotation (i.e. the R to E transition), and another having previously had an RM gradient imposed across it  (i.e. the I to E transition).

\subsubsection{PKS B0611-74}

\citet{Murphy2010} measure fractional polarisations of 10.9\%, 11.0\%, and 9.9\% at 20, 8, and 5 GHz in 2009. The latter two measurements are broadly consistent with the 2012 epoch data from A16, but not consistent with the 2017 epoch data presented in this work. 

The source is well-fit by an IEE model in 2012 and an RII model in 2017, though an IIE-type model represents a plausible alternative in the former case. We associate the $j=2$ component in 2012 with the $j=3$ component in 2017 to form component C10 in Table \ref{table:compAssociationResults} and Figure \ref{fig:pvdplots}, owing to the similarity in the associated $p_0$, $\psi_0$, and RM values. We form component C11 from the $j=3$ component in 2012 and the $j=1$ component in 2017, owing to their similarity in the  $p_0$ and $\psi_0$ values. Component C12 is formed from the remaining $j=1$ and $j=2$ components in 2012 and 2017 respectively. The important changes between epochs are thus a large drop in the detectable intrinsic fractional polarisation of component C12, an increase of 30 rad m$^{-2}$ in the RM of component C11, and  changes to the type and magnitude of depolarisation of all of the components --- most notably a $1175$ rad m$^{-2}$ drop in the Faraday dispersion of C12. Our cross-epoch component associations suggest that an E-type component changes to an I-type component, and that an E-type component transitions to an R-type component. One interpretation of this is that the two components formerly illuminated a turbulent foreground Faraday screen from behind, but now illuminate relatively less-structured parts of the screen, which induce only an RM gradient and Faraday rotation. 

\subsubsection{PKS B1039-47}\label{sec-nuthaVLBI}

This source is an FSRQ located at $z=2.59$ (OS12). It is unresolved at 0.15 arcsec (1.2 kpc, \citealp{Chhetri2013}), but resolves into three knots in Stokes $I$ that span $\sim20$ mas (160 pc) along a line oriented at $\sim$-70 degrees \citep{Ojha2004}. OS12 fit a RRR-type model (in our parlance) to its polarisation spectrum, with component RMs of $-13$, $-30$, and $+68$ rad m$^{-2}$ respectively. They suggest that these spectral components directly correspond to the emission knots revealed by VLBI. \citet{Murphy2010} measure fractional polarisations of 1.8\%, 3.8\%, and 3.9\% at 20, 8 and 5 GHz in 2009 (respectively), the latter of which is broadly consistent with an extrapolation of OS12's best-fit model.  

We find that an MEE model is preferred in 2012, and an RMI model is preferred in 2017, though there are a large number of plausible alternatives to the latter. We associate the $j=1$ and $j=2$ components in 2012 and 2017 (respectively) to form C13, since their RM value is identical within the uncertainties. We then associate the $j=3$ components with one another across epochs to produce component C14, which results in insignificant changes in the $p_0$, $\psi_0$, RM, and $\sigma_\text{RM}$ parameters. The remaining $j=2$ and $j=1$ components then form C15. The most significant changes in the source are then a 3\% increase in $p_0$ and a 62 rad m$^{-2}$ increase (n.b. 800 rad m$^{-2}$ in the source frame) in RM for component C15, and changes in the Faraday dispersion properties of all components. For two of the components, their type changes from external depolarisation to internal depolarisation and pure Faraday rotation respectively, which may come about due to shifting illumination of a foreground Faraday screen, as previously described.

\subsubsection{PKS B1610-771}\label{sec-tingayEvolvingSource}

This source is an FSRQ (e.g. \citealp{Healey2007}) located at $z=1.71$ \citep{HM1980}. It is unresolved at 0.15 arcsec (1.3 kpc; \citealp{GH2000,Chhetri2013}) but resolves into several bright Stokes $I$ knots on milliarcsecond scales at 2.3 GHz and 8.4 GHz. The jet is initially aligned at a position angle (p.a.) of $\sim$30 degrees \citep{Tingay2002}, deviates to a p.a. of -70 degrees at scales of tens of parsecs, but thereafter returns to a $\sim$-30 degree p.a. out to $\sim130$ pc (projected; \citealp{Ojha2010}). \citet{Tingay2002} report a separation rate of 0.2 mas yr$^{-1}$ between the most compact knots, corresponding to an apparent transverse speed of $\sim9$h$^{-1}c$ (where $c$ is the speed of light, and the Hubble constant H$_0=100h$ km s$^{-1}$ Mpc$^{-1}$). In unrelated observations, \citet{Tingay2003} established the existence of a systematic decrease in the (spatially-integrated) polarised fraction over 3.5 years starting in 1996 (from 3--6\% down to 0--3\% at each of 8.6, 4.8, 2.5, and 1.4 GHz; see Figure 2 of that work; see also Section \ref{sec-indivFirstSource}). \citet{Murphy2010} measure fractional polarisations of 1.6\%, 2.1\%, and 2.3\% at 20 GHz, 8 GHz, and 5 GHz in 2009 --- consistent with the decreasing trend established by \citet{Tingay2003}, but considerably lower than what would be inferred by extrapolation of the OS12 best-fit model, or the 2017 measurements presented here (i.e. $\sim$3.5\% at 8 GHz; $\sim$4.5\% at 5 GHz). 

The source is well-fit by an RMI model in 2012 and an II model in 2017 --- in both cases slightly preferred over two other plausible model types. Associating the $j=2$ components with each other across epochs to yield C16 results in close matches or insignificant differences in $p_0$, $\psi_0$, RM, and $\Delta\phi$. The $j=1$ component from 2017 does not align more closely with either the $j=1$ or $j=3$ components from 2012, so we leave these unassociated. The M to I -type component transition could, for example, arise if an emission component with mixed synchrotron-emitting and Faraday-rotating plasmas emerged from behind a turbulent Faraday screen in the foreground. Alternatively, the transition could be caused by hydrodynamic sheer acting on an emitting region consisting of \emph{turbulent} mixed synchrotron-emitting and Faraday-rotating plasmas (as M-type components can describe (e.g. \citealp{Sokoloff1998,Anderson2016,OSullivan2017}), acting to stretch the magnetic field lines and make the magnetised structure of the region more uniform.

\subsubsection{PKS B1903-802}\label{sec-indivLastSource}

This source is an FSRQ located at $z=0.50$ (OS12). It is unresolved at 0.15 arcsec (0.9 kpc; \citealp{Chhetri2013}), but resolves into two knots in Stokes $I$ on scales less than 5 mas (30 pc) at 8.4 GHz \citep{Ojha2005}. OS12 fit the source with a single E-type component, having $\text{RM}=+18$ rad m$^{-2}$ and $\sigma_\text{RM}\approx5$ rad m$^{-2}$. However, \citet{Murphy2010} measure fractional polarisations of 3.4\%, 5.7\%, and 3.0\% at 20 GHz, 8 GHz, and 5 GHz during a one-month period in 2009, implying the existence of Faraday complexity not seen in the OS12 frequency range. This is supported by \citet{Tingay2003}, who measure complex frequency-dependent changes in the polarised fraction (but not intrinsic angle, which remains remarkably constant) at 8.6, 4.8, 2.5, and 1.4 GHz over a period of $\sim3.5$ years (see their Figure 2). We \emph{do} capture similar complexity in our 2017 data, though the fractional polarisations differ from the Murphy et al. data, adding further evidence for variability. 

The source is well-fit in 2012 by an IE type model (though with plausible alternatives), and reasonably well-fit by an MII model in 2017. Associating the $j=1$ component across epochs to yield component C17 leads to small or insignificant changes in all best-fit parameter values. We associate the $j=2$ and $j=3$ components from 2012 and 2017 respectively as component C18, which produces little net change in RM. There is no other component in the 2012 best fit model to associate with the remaining $j=2$ component in 2017. The E to I -type component transition could arise as described previously, while the I to M -type component transition could arise, for example, as a result of the onset of turbulence in an emission region characterised by mixed synchrotron-emitting and Faraday-rotating plasmas.

\section{Discussion}\label{sec-discussion} 
This is the first study of variability in the \emph{densely-sampled} (in frequency) multi-GHz-band polarisation spectra of blazars. For this pilot survey, we obtained the broadband polarisation spectra for nine sources over multi-GHz-bandwidths in two epochs separated by $\sim5$ years. We fit multi-component Faraday rotation and depolarisation models to these data, aiming to: 

\begin{itemize}
\item characterise spectropolarimetric variability in the blazar class, using densely-sampled broadband polarisation data for the first time 
\item search for temporal changes in the Faraday depth structure of the sources
\item determine the most plausible physical origin of any changes observed 
\item establish whether multi-epoch broadband polarimetry might be used as a (limited) proxy or complement for VLBI polarimetry, in the manner proposed by L11, OS12, and A16.
\item identify consequences and opportunities for next-generation broadband radio surveys
\end{itemize}

We now discuss our results in light of these aims.

\subsection{Veracity of the observations and limitations of the modelling}\label{sec-veracity}

We detected spectropolarimetric changes of varying magnitude in all of the polarised sources in our sample. The data for PKS B0545-649 and PKS B1610-711 are notable, because they changed only a little (see Figs. \ref{fig:04_comparo} and \ref{fig:1610_comparo}), and because the sources were selected from A16 and OS12 respectively. This demonstrates that the calibration of all three studies is reliable, and that the larger spectral changes observed in the remainder of the sample are real. Nevertheless, we contend that even the small changes seen in PKS B1610-711 are also real, based on our analysis in Section \ref{sec-bfmodelcomparo}.  

The best-fit polarisation models are well constrained (see Appendix \ref{sec-appendA}, which is available in the online version of this paper) but contain up to 15 fitted parameters, leaving numerous model types of comparable complexity (but less generality) untested (e.g. see \citealp{Sokoloff1998}). Moreover, the model types we \emph{did} test do not account for variation in spectral index or optical depth among emitting regions. This is probably unavoidable if prior information about the source is not available, but might be ameliorated in joint VLBI experiments (see Section \ref{sec-discussion-fauxVLBI}). Despite this, the intrinsic angle of polarised blazar emission either does not vary significantly with optical depth (e.g. \citealp{ZT2004}), or otherwise originates mostly from towards optically thin regions (e.g. \citealp{Gabuzda2015c,Wardle2018}). When the latter condition does not hold, the most acute effects of optical depth on polarisation are likely suppressed precisely because of the high optical depths \citep{Cobb1993,Wardle2018} and frequencies \citep{Porth2011} at which they are generated. On the other hand, VLBI studies show that blazar jets are often dominated by a handful of bright emission components with differing polarised emission, Faraday rotation, and Faraday depolarisation properties (e.g. \citealp{Hovatta2012}) --- the existence of which \emph{must} give rise to frequency-dependent interference effects in the integrated polarisation signal (e.g. \citealp{Slysh1965,Burn1966,GW1966,Conway1974}). We therefore consider the range of model types tested here to be appropriate for the aims of our study.

For the OS12 sources, we used fitting results for the 16cm band only in our multi-epoch analysis (see Section \ref{sec-modelling}; also column 5 of Table \ref{tab:SourceDat}). Unsurprisingly, the resulting best-fit models do not describe the 2017 epoch 6/3 cm band data well (see Figs. \ref{fig:0454_comparo}, \ref{fig:1039_comparo}--\ref{fig:1903_comparo}). From columns 11 and 12 of Table \ref{table:BFModResults}, it is clear that these discrepancies are the result of emission components that depolarise strongly in the higher frequency bands, highlighting the practical need for dense, broadband radio frequency coverage to accurately reconstruct the Faraday depth structure of blazars. The upcoming Q and U Observations at Cm bands and Kilometre baselines with the ATCA (QUOCKA) survey (Heald et al., \emph{in prep.}) will investigate degeneracies of this type in detail.

\subsection{The nature of the observed variability}\label{sec-siteof}

Having eliminated calibration or observational effects as the cause of the spectral changes, it must be that either (1) the configuration of the sources themselves has changed, or (2) the interstellar scintillation has caused time-variable focusing and defocusing of the polarised substructure in the radio jets (e.g. \citealp{Rickett2001,dBM2015}), which can generate spectropolarimetric variability (e.g. \citealp{Kedziora-Chudczer2006}). ISS is difficult to distinguish from intrinsic variability without fully time-resolved data (e.g. see \citealp{Bignall2015} and references therein), and can coexist with it \citep{Koay2018}. Nevertheless, ISS does not easily explain our results: Sources located more than $\sim25^\circ$ from the Galactic plane (3/4 of our polarised sources) and with line-of-sight H$\alpha$ intensities less than $\sim$ a few Rayleighs (7/8 of our polarised sources; see column 9 of Table \ref{tab:SourceDat}) are typically not strongly scattered \citep{Pushkarev2015}, and experience associated flux density modulations of typically less than a few per cent \citep{Heeschen1984,Quirrenbach1992,RLG2006,Lovell2008}. This is smaller than the changes observed in our integrated \emph{fractional} polarisation spectra, which should undergo less modulation than Stokes $I$ in any case. ISS modulation also has a distinct frequency and time dependence. At the mid-galactic latitudes inhabited by most of ours sources (see column 8 of Table \ref{tab:SourceDat}), its magnitude peaks between 4 and 6 GHz ($\lambda^2$ between 0.0024 and 0.0056 m$^2$), and drops to less than 50\% of this value at 1.4 GHz and 10 GHz (i.e. $\lambda^2$ of 0.0008 and 0.046 m$^2$). Thus, changes due to ISS should `spike' in this narrow $\lambda^2$ window; which we see little evidence of (though see perhaps PKS B0517-726 and PKS B1903-802; see Figs. \ref{fig:13_comparo} and \ref{fig:1903_comparo}). The characteristic timescale of ISS is $\sim$hours to days (\citealp{RLG2006,Gabanyi2007}, and references therein), and over multi-year timescales, intrinsic effects tend to dominate the observed variability (e.g. \citealp{Lazio2001,RLG2006,Mooley2016}). We therefore claim that the spectral variability most likely reflects intrinsic changes in the sources themselves.


\subsection{Comparing our results to the known properties of blazars}\label{sec-bfmodeldisco}

For the concluding claim of the previous section to be true, our spectral flux densities must be dominated by emission components lying between 1--10s of parsecs from the central SMBH. Thus, we now compare our best-fit parameter values in $\psi_0$, RM, and $\sigma_{\text{RM}}$ and $\Delta\phi$ to the known properties of blazars at these scales. 

\begin{figure}[htpb]
\centering
\includegraphics[width=0.45\textwidth]{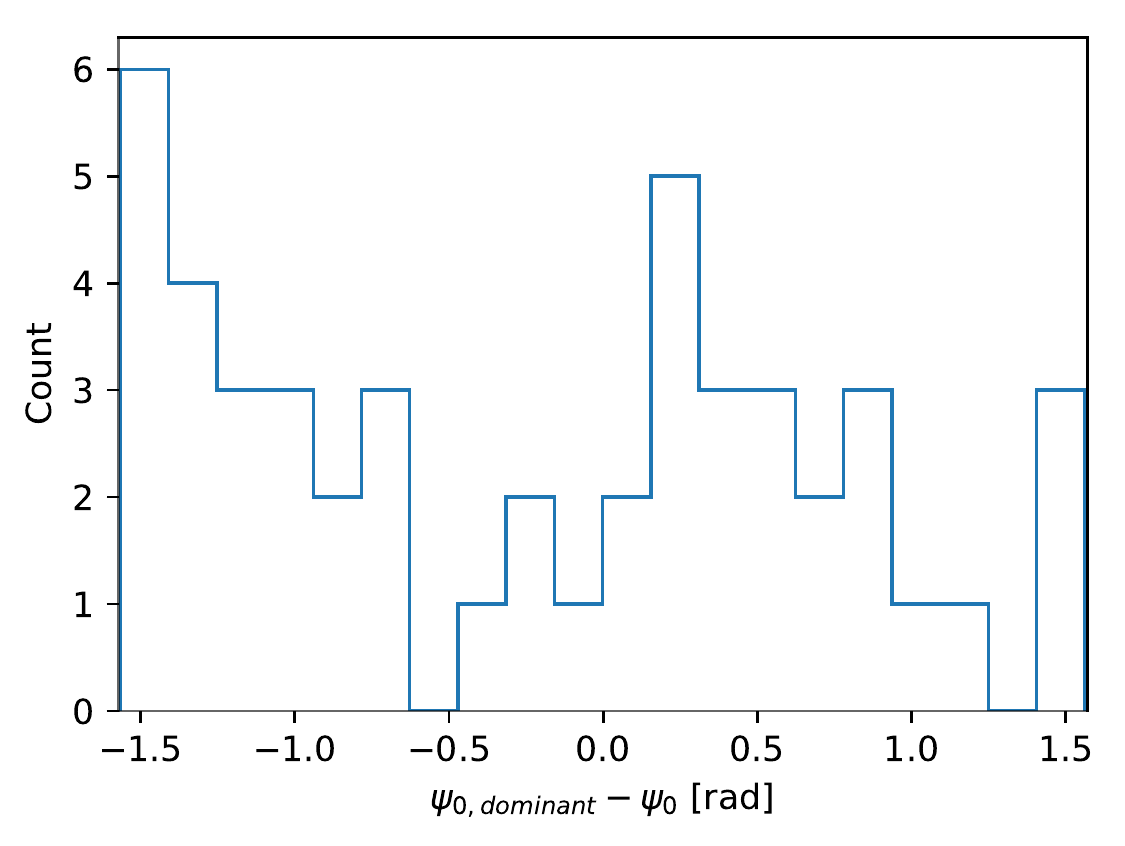}
\caption{Distribution of intrinsic polarisation angle differences for emission components in our best fit models. The values were calculated on a per-source basis (from data presented in column 9 of Table \ref {table:BFModResults}), wrapped to the interval [-$\pi/2$,$\pi/2$), then accumulated for all sources in both epochs.}
\label{fig:angleshist}
\end{figure}

We consider $\psi_0$ first. The upper righthand panel of Figure \ref{fig:pvdplots} plots the histograms of $\psi_{0, 2012}$ and $\psi_{0, 2017}$, and changes in $\psi_0$ between epochs for associated components. As expected for physically unrelated sources, the $\psi_0$ values are evenly distributed over [-$\pi$/2,+$\pi$/2) radians. However, the distribution of the difference in $\psi_0$ between emission components calculated for each source is clearly bimodal, with peaks separated by $\sim\pi/2$ radians (see Figure \ref{fig:angleshist}). Similar results were described by A16 and \citet{OSullivan2017}; it is consistent with the bi-modal distribution of jet-axis-aligned and jet-axis-perpendicular intrinsic polarisation angles in blazars (e.g. \citealp{Attridge1999,Lyutikov2005,Pushkarev2005}), and we interpret it as further evidence to suggest that our modelling recovers physical parameter values accurately. If verified by robust comparison to VLBI data, it suggests we can predict the projected jet orientation (modulo $\pi/2$ radians) of spatially unresolved sources, which might be exploited (for example) to probe cosmic radio jet alignment effects (e.g. \citealp{TJ2016}) in upcoming all-sky polarimetric surveys in a statistical manner. 

A histogram of our best-fit RM values is plotted in Figure \ref{fig:pvdplots}, along with a histogram of absolute RMs in blazar jet components located at de-projected distances of up to 100 parsecs from the central SMBH (derived from RM data plotted in Figure 4 of \citealp{Hovatta2012}, calculated between 8 and 15 GHz). Evidently, our absolute RM distribution  matches that measured by \citet{Hovatta2012} reasonably closely. Our median RM value (59 rad m$^{-2}$; Section \ref{sec-bfmodelsselect}) is $\sim$40\% lower than that reported by \citet{Hovatta2012} (102 rad m$^{-2}$) for isolated jet components in the observers frame. VLBI-derived RMs tend to depend on the observing frequency (which, we point out, highlights limitations associated with the restricted VLBI bandwidths), and the \citet{Hovatta2012} RM values were calculated between 8 and 15 GHz. However, this difference might also indicate that we are sensitive to emission on larger scales in some proportion of sources, which is resolved out by the VLBA. Comparison of our results with low frequency (near 1.4 GHz) VLBI observations would be valuable here, since they tend to probe emission at the tens-of-parsecs scale by virtue of extended $uv$ coverage, and would also facilitate comparison at matched frequencies. Few such studies have been published to date, but those that have tend to reveal RM magnitudes of $\sim$10s of rad m$^{-2}$ and below \citep{HG2008,Coughlan2010,Croke2010,Gabuzda2012,Gabuzda2014,MG2017} --- the lower end of our absolute RM distribution encompasses this range. 

\begin{figure}[htpb]
\centering
\includegraphics[width=0.45\textwidth]{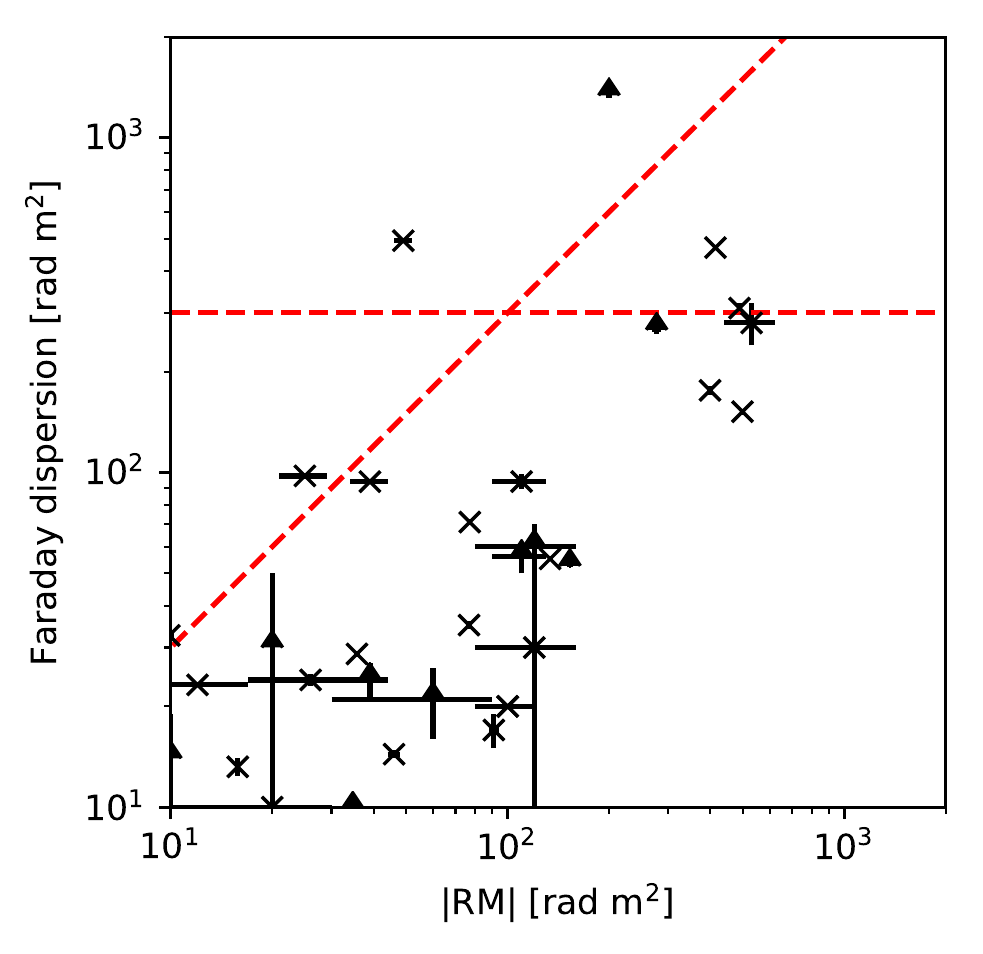}
\caption{Plots of $\sigma_{\text{RM}}$ (triangles) and $\Delta\phi$ (crosses) versus the magnitude of the RM for components in which depolarisation of the given type is present. We include data for both the OS12 and A16 sources; the differences in frequency coverage between these subsamples should be borne in mind. The diagonal red dashed line indicates where $\sigma_{\text{RM}}$ and $\Delta\phi$ is three times the RM (see main text). The red horizontal line indicates where the Faraday dispersion is equal to 300 rad m$^{-2}$ (again, see discussion in main text).}
\label{fig:dispsVsRMs}
\end{figure}

Considering now our results for $\sigma_{\text{RM}}$ and $\Delta\phi$, \citet{Hovatta2012} identified two scenarios consistent with their measured relationships between depolarisation, Faraday dispersion, and RMs in isolated jet components in their blazar sample (see their Section 4.2 and Figure 7). That is, the jet components shine through either (1) a homogeneous, turbulent, magnetised plasma in the immediate foreground (see e.g. \citealp{Burn1966}) in which $\sigma_\text{RM}\sim\sqrt{10}$RM $\approx3$RM, or (2) an inhomogeneous turbulent magnetised plasma for which RM$=0$ rad m$^{-2}$ and $\sigma_\text{RM}=300$ rad m$^{-2}$, intercepting roughly ten independent turbulent `cells' during transit (see e.g. \citealp{Tribble1991}). Our data are broadly consistent with these scenarios too: Plotting both $\sigma_{\text{RM}}$ (triangles) and $\Delta\phi$ (crosses) vs. RM in Figure \ref{fig:dispsVsRMs}, we see that the relationships $\sigma_\text{RM}=3$RM and $\sigma_\text{RM}=300$ rad m$^{-2}$ (or the $\Delta\phi$ equivalents) pass through our best fit data. While we also have an abundance of points with comparatively low Faraday-dispersion-to-RM ratios, we note that (a) these are mainly associated with the OS12 sources, which are insensitive to Faraday dispersions greater in magnitude than 85  rad m$^{-2}$ , and (b) Hovatta et al. excluded components with low levels of depolarisation (small $\sigma_\text{RM}$ and $\Delta\phi$ values) from the analysis in which they identified the aforementioned relationships. 

Finally, we consider our cross-epoch component associations, and whether the associated parameter value changes comport with results from the single dish monitoring and VLBI literature. In the fractional polarisation of spatially-unresolved emission components in blazars, changes of between a few and 10\% are commonly observed over multi-year timescales, as are intrinsic angle rotations of a few hundred degrees (e.g. \citealp{Aller2017}, and refs therein). These are similar to the results presented in Figure 10 and columns 4 and 6 of Table 4. In VLBI experiments, changes in RM of up to thousands of rad m$^{-2}$ are observed over multi-year timescales in jet components (e.g. \citealp{ZT2001,Asada2008,Gomez2011,Hovatta2012}), but conversely, those with stable RMs are also observed (\citealp{Gomez2011,Hovatta2012}). We claim to detect changes in RM of typically tens of rad m$^{-2}$ and sometimes up to several hundreds of rad m$^{-2}$, but also occasionally identify components with stable RM values (see sections \ref{sec-indivFirstSource}--\ref{sec-indivLastSource}, Figure 10, and column 8 of Table 4). Our reported RM changes are typically smaller than those measured with VLBI, but the RM uncertainties on the latter are often large due to limited $\lambda^2$ coverage. It is interesting then to speculate that broadband observations could provide a more sensitive diagnostic of time-variable RMs in blazars than afforded by VLBI. Temporal variability in RM gradients across emission components have also been described in the literature. \citet{Mahmud2009}, for example, describe an RM gradient that changes sign over a period of a few years, and changes its span in RM from several tens of rad m$^{-2}$ to almost 1000 rad m$^{-2}$ in the process. Monotonic RM gradients can map to Faraday thick structures in the FDF of spatially unresolved sources (e.g. \citealp{Schnitzeler2015,Anderson2016}), similar to the E- and I--type components in the modelling we undertake in this work. The changes we report in $\sigma_\text{RM}$ and $\Delta\phi$ for our cross-epoch component associations are comparable in magnitude to the \citet{Mahmud2009} results (Figure 10 and columns 10 and 12 of Table 4). 

In summary then, our fitted angles, RMs, Faraday dispersions, and the changes in such, are broadly consistent with the known properties of quasar environments on scales of parsecs to tens of parsecs, established through VLBI polarimetry and single-dish monitoring. While this is suggestive rather than definitive, we contend that the spectral variability that we have observed reflects changes in the emissivity, polarised fraction, or polarisation angle of components in the blazar jets, the RM and Faraday dispersion induced by an evolving line of sight through the back-illuminated foreground environment, or some combination of these. Given this conclusion, we now consider possible scientific applications and implications of our findings.

\subsection{A probe of evolving blazar jets}\label{sec-discussion-probeofblazars} 

\subsubsection{Enhancing time-resolved monitoring of integrated emission}\label{sec-discussion-timeresolve} 

Single dish monitoring studies have tracked changes in polarisation at multiple frequencies over a number of decades (see Section 1), but the sparseness of the frequency sampling does not capture the subtle frequency-dependent structure that can be generated by Faraday effects. We have shown that such structure is present in blazar spectra, and that it can evolve with time. Since the spectral interference effects in question can be strongly affected by the structure and topology of magnetic fields in the emitting regions, and different models of blazar emission predict different temporal evolution of such, time-resolved broadband polarimetry could in principle help distinguish between competing models such as the shock-in-jet model, or domains of magnetic reconnection (see Section \ref{sec-intro}). On this basis alone, we suggest that high-cadence broadband polarimetry represents a valuable and natural direction in which blazar monitoring experiments can progress.

\subsubsection{A complement or limited proxy for VLBI polarimetry}\label{sec-discussion-fauxVLBI} 

Multi-epoch, multi-frequency, VLBI polarimetry is a mature and powerful technique for studying the magnetoionic structure of blazar jets and their environments (see Section 1). However, these observations are also time-consuming, have comparatively demanding source selection requirements, and operate within comparatively limited windows of frequency and spatial scale-size sensitivity (e.g. \citealp{TZ2010,Pudritz2012}). In principle, broadband spectra encode the emission properties of components irrespective of spatial resolution --- their prominence in the spectra depends only on their brightness and the rate at which they depolarise relative to the $\lambda^2$ coverage of the observations. With greater temporal sampling density, we suggest that global changes in the properties and configuration of a system might be precisely measured through modelling of its dynamic polarisation spectrum. In particular, trends in modelled quantities might be used to isolate changes occuring in specific emission components more concretely. The following are suggested as scientific use cases:\\

\emph{Monitoring/triggering}: Observations of bright polarised sources require only short integration times to achieve a high S/N ratio. For example, the Australia Telescope Extreme Scattering Events (ATESE) survey \citep{Bannister2016} uses the ATCA to measure the (Stokes $I$) spectra of $\sim1000$ compact radio sources each month over a 24 hour period with high signal-to-noise. Similar observational strategies could be used to observe the polarisation spectrum of samples of AGN or microblazars, and follow-up observations could be triggered when interesting phases of activity occur --- for example, the emergence of a bright polarised component, or rapid increases or decreases in depolarisation. \\

\emph{Intensive targeted studies}: Notwithstanding their evident power, VLBI studies generally make use of sparsely sampled frequency bands, which can provide a misleading view of magneto-ionic structure along a sight-line (e.g. see \citealp{BdB2005,OSullivan2012,Anderson2016b}). Broadband polarimetry is sensitive to changes in the magnetoionic configuration of a source, but with limited capacity to interpret this in context. We suggest the information provided by each is complementary and can be usefully combined: Single- or multi-epoch VLBI observations can be used to inform spectropolarimetric modelling, including on the number of bright emission components in a source, and their spectral index, optical thickness, polarisation angle, polarised intensity, and apparent motion. Subsequently, high-cadence broadband polarimetry might be used to track subtle changes in the properties of these components as they move and evolve. The source PKS B1610-771 (Section \ref{sec-tingayEvolvingSource}) --- in which VLBI observations have detected multiple emission components in relative motion to one another, and in which we have detected and modelled changes in its integrated polarisation spectrum --- serves to illustrate this possible synergy. Were we to continue to track changes in the observable Faraday depth and dispersion of this source with high cadence, we might perhaps map out the properties of the circum-jet environment (e.g. \citealp{Asada2008,Gomez2011,Lico2017,Wardle2018}). Such analysis might be particularly useful for micro-blazars, whose outburst and ejection timescales are short, and whose propagating emission structure relatively simple (e.g. \citealp{Egron2017}).\\

\emph{Statistical characterisation of jetted sources}: With the advent of powerful new radio survey telescopes such as the Australian Square Kilometre Array Pathfinder (ASKAP; \citealp{Johnston2007,McConnell2016}), the Murchison Wide-field Array (MWA; \citealp{Tingay2013}), the Low-Frequency Array (LOFAR; \citealp{vh2013}), the Karl G. Jansky Very Large Array (VLA; \citealp{Perley2011}), and the Karoo Array Telescope (MeerKAT; \citealp{Jonas2009}), it will be possible to observe large areas of sky repeatedly with broad observing bands (e.g.  the Variables and Slow Transients (VAST) survey on ASKAP; \citealp{Murphy2013}). Thus, time-resolved broadband polarimetry may be able to track changes in the global magnetoionic structure of large samples of radio-loud AGN, providing a powerful statistical dimension to pursue the types of questions and analysis we have already described.


\subsection{Implications for broadband polarisation surveys}\label{sec-discussion-ramify} 

We now briefly comment on implications for forthcoming polarimetric surveys. RM grid surveys seek to exploit ensembles of stable, linearly polarised background sources as probes of foreground material via statistical analysis of their modified Faraday rotation measure. Sources of the type considered in this paper (i.e. those in which the local source environment contributes much of the observed Faraday depth structure) should be excluded from such, or otherwise appropriately considered in the analysis. Around $\sim$10-20\% of sources in an NVSS-like survey (i.e. $\sim$GHz-frequency, $\sim$arcminute resolution, mJy flux density limit) will have a flat spectrum \citep{Tucci2004}, and might therefore be considered blazars for the purpose of this argument. Of these, it is not known what proportion will show the variability effects presented here. However, \cite{Tingay2003} used the ATCA to monitor a sample of 167 flat spectrum sources for 3.5 years at five discrete frequencies between 1 and 9 GHz, and found that approximately one in three varied in total intensity by $\sim$10\%, while one source in 20 varied by $30$\% or more. Assuming that variability in the polarisation emission largely tracks that of the total intensity, then perhaps $\sim$5\% of RM grid sources could show effects similar to those described in this work. This should be borne in mind, but will not significantly impact RM grid science. More important is that in the era of all-sky radio surveys, science will be pursued by combining polarimetric data from different bands and epochs to gain enhanced $\lambda^2$ coverage --- e.g. from the VLASS \citep{VSSG2015} and POSSUM \citep{Gaensler2010} surveys. The caution required here is obvious.

\section{Summary and conclusion}\label{sec-conclude} 

We have shown how the polarisation spectrum of blazars can vary in time, using dual-epoch observations over a densely-sampled 1.3--9.9 GHz band. We were able to model the blazar spectra with general multi-component Faraday rotation and depolarisation models. We determined that the observed spectral changes could not be attributed to calibration or observational effects, and were unlikely to be predominantly caused by interstellar scintillation. We showed that the fitted intrinsic polarisation angles, RMs and Faraday dispersions of components in our samples were broadly consistent with the known properties of blazars on scales of parsecs to $\sim$ several tens of parsecs. On this basis, and by a process of elimination, we concluded that the observed spectra changes most likely originate due to evolution in the parsec to decaparsec scale structure of the sources. If our interpretation of the data are correct, our results pave the way for the effect to be exploited as a novel probe of the properties of AGN and micro-quasars on angular scales that might otherwise be inaccessible.

We are building on this work by analysing archival multi-epoch data from the ATCA Calibrator Database, consisting of hundreds of sources observed over broad bands at three or more epochs, and with extensive ancillary data available to aid interpretation.   

\section{Acknowledgements}\label{sec-acknowledgements}

We thank the anonymous referee for their time, and for constructing a very useful report. Our paper has benefitted from this significantly. We also thank Wasim Raja for undertaking the observations for this work. The Australia Telescope Compact Array is part of the Australia Telescope National Facility which is funded by the Commonwealth of Australia for operation as a National Facility managed by CSIRO. The Dunlap Institute is funded through an endowment established by the David Dunlap family and the University of Toronto. The University of Toronto operates on the traditional land of the Huron-Wendat, the Seneca, and most recently, the Mississaugas of the Credit River; B. M. G. is grateful to have the opportunity to work on this land. B. M. G. acknowledges the support of the Natural Sciences and Engineering Research Council of Canada (NSERC) through grant RGPIN-2015-05948, and of the Canada Research Chairs program.


\def\latex{\LaTeX{}}
\def\azh{Astronomicheskii Zhurnal}                   
\def\aj{AJ}                   
\def\araa{ARA\&A}             
\def\apj{ApJ}                 
\def\apjl{ApJL}                
\def\apjs{ApJS}               
\def\apss{Ap\&SS}             
\def\aap{A\&A}                
\def\aapr{A\&A~Rev.}          
\def\aaps{A\&AS}              
\def\mnras{MNRAS}             
\def\nat{Nature}              
\def\pasp{PASP}               
\def\pasa{PASA}
\def\pasj{PASJ} 
\def\nar{NewAR}
\def\skytel{Sky \& Telescope}              
\def\ssr{Space Sci. Rev.}  
\def\iaucirc{IAU circ.}

\bibliographystyle{apj}
\bibliography{./bibliography}

\appendix

\section{Marginal posteriors for the best-fit model types}\label{sec-appendA}

In this section we provide the marginal and conditional posterior distributions for the best-fit model type for the complex polarisation spectrum for an example source and epoch. The modelling and model selection is described in Section \ref{sec-analysis}. The plots for all sources and epochs are provided in the supplementary online materials for the journal version of the paper.

\begin{figure*}[htpb]
\centering
\includegraphics[width=\textwidth]{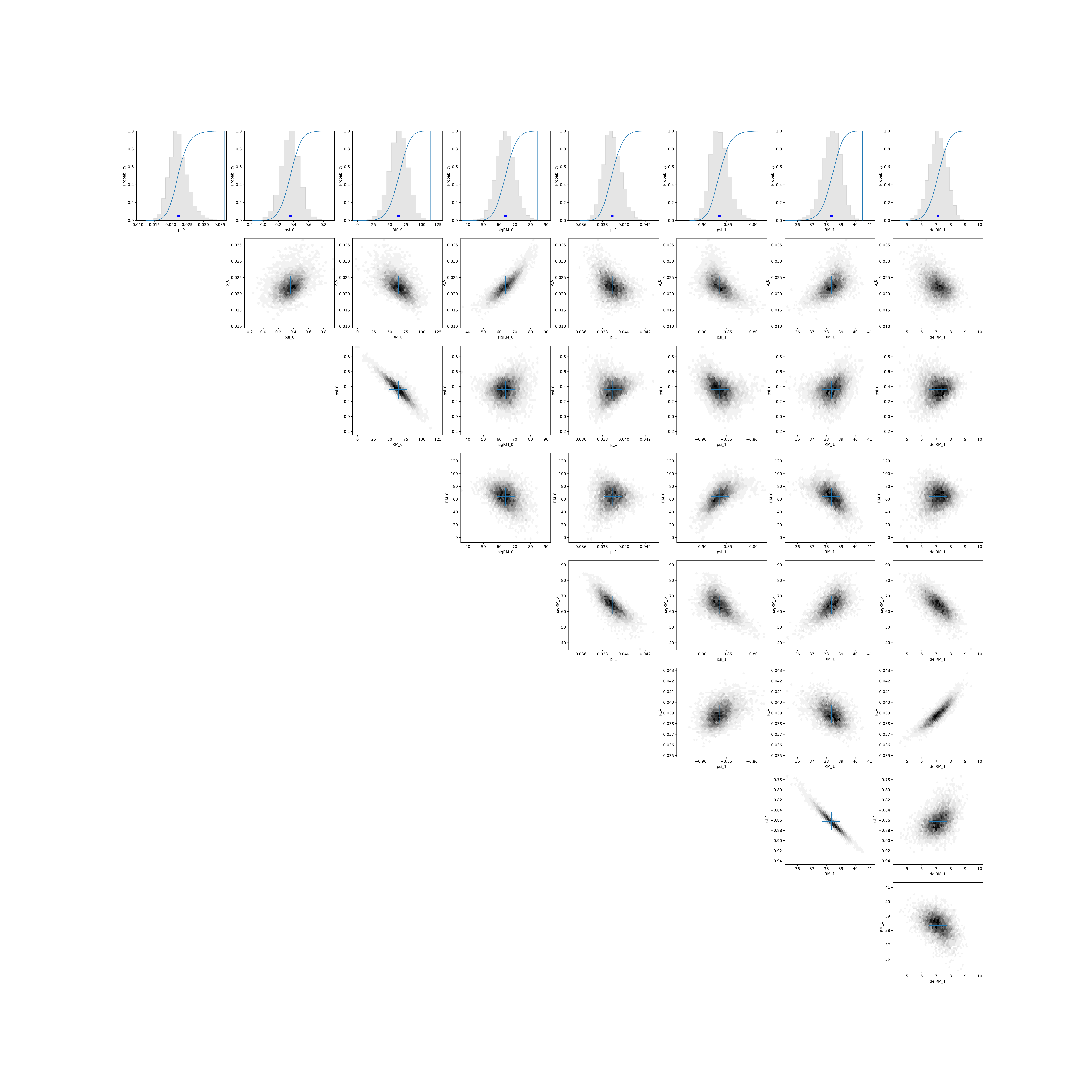}
\caption{Marginal posterior distributions over parameters in the best-fit model type for the source PKS B0454-810 in the 2012 epoch.}
\label{fig:0454_arch_marg}
\end{figure*}

\end{document}